# Compensating fields in the Landau local theory and phenomenological description of the electron-phonon interaction.

Alexander Ya. Braginsky

Southern Federal University, Research Institute of Physics, Rostov-on-Don

e-mail: **a.braginsky@mail.ru**

**Abstract.**

We study inhomogeneous states with local translation symmetry, described by the order parameter (OP) with local transformation properties $\vec{k} = \vec{k}(\vec{X})$. It is shown that in the OP extended derivative we should consider compensating field which is equivalent to the distortion tensor. The definition of the phonon potential as a 4-distortion tensor is given and equations of motion of a particle in a phonon field are obtained. We construct the Ginzburg-Landau potential, which clearly describes the electron-phonon interaction and gives the correct definition of the quantum of magnetic flux without doubling the phenomenological charge interaction. The description of the states with the d-symmetry of the wave superconducting gap by the low-symmetric solutions is suggested, in which the inversion of space-time is not equivalent to the OP complex conjugation. A comparative analysis of the nonlinear Landau theory with the compensating field distortion and the Kadić-Edelen linear gauge theory of dislocations is performed. It was demonstrated that the minimal interaction is a consequence of local wave properties of OP, which is consistent with the basic principles of quantum mechanics.

PACS number(s): 64.60.Bb, **74.20.-z.**

## 1. Introduction.

Gauge models in the Landau theory with minimal interaction were introduced in the pioneer paper by Ginzburg and Landau [1], when describing the phase transitions in the superconducting state. Later de Gennes [2] using a calibration model described the screening of the stress field and dislocations in the deformed SmA. In both cases, the model with the extended derivatives in the Landau theory were taken from the gauge field theory [3]. In this paper we show that the minimal interaction between the compensating field and OP, which is given by the extended derivative, bound to appear in the Landau local theory. It is a consequence of translational invariance of the local Landau potential, so in [1,2] there is no need to appeal to the gauge field theory.

The transformation properties of OP with respect to the subgroup of spatial translations are characterized by the vector $\vec{k}$ of the irreducible representation (IR). The OP local transformation properties - the dependence of $\vec{k}$ vector on the coordinate [4], necessarily lead to the appearance of additional compensating fields in the OP extended derivative. At the same time translationally invariant combinations of the compensating fields correspond to the observed physical quantities such as magnetic induction $\vec{B}$ and density of dislocations $\rho_{ij}$. The physical condition of such a model is characterized by a pair: OP and compensating field.

Let's revise the basic concepts of the Landau theory to determine the local Landau potential and explain what is meant by the dependence of the vector $\vec{k}$ from the coordinates. As it is known [5], the Landau theory deals with two fundamental concepts, the first being the OP, which describes the change in the symmetry of the state density in case of the phase transition.

To determine the OP the change in the state density $\delta\rho(\vec{x})$ must be represented as an expansion of the basis functions of the IR space group $G_0$ of the high-symmetric phase symmetry:

$$\delta\rho(\vec{x}) = \sum_n \iiint_\Omega c_{\vec{k}nl} \varphi_{\vec{k}nl}(\vec{x}) d\vec{k} . \qquad (1)$$

Here $\varphi_{\vec{k}nl}(\vec{x})$ is a function of the IR basis of the group $G_0$, they are characterized by a star of the vector $\vec{k}$ and the number $n$ of the IR star of the vector $\vec{k}$, as well as a component $l$ of the IR. The expansion coefficients of the function $\delta\rho(\vec{x})$ into $\varphi_{\vec{k}nl}(\vec{x})$ - $c_{\vec{k}nl}$ define the symmetry $\delta\rho(\vec{x})$. In the Landau theory [5] it is proved that in the phase transitions of the second order in (1) it is sufficient to retain only those functions $\varphi_{\vec{k}nl}(\vec{x})$ that are converted by one IR and have identical indexes $\vec{k}_0$ and $n_0$. The coefficients $c_{\vec{k}_0 n_0 l}$ are the components of the OP, hereinafter, for short, denoted as $c_{\vec{k}_0 n_0 l} \equiv \eta_l$.

The second fundamental concept in the theory of phase transitions is a nonequilibrium Landau potential $F$. It is functional $\varphi_{\vec{k}_0 n_0 l}(\vec{x})$ and, therefore, a function of the component $\eta_l$ of the OP. Nonequilibrium potential $F$ is an invariant of the group $G_0$, so it depends only on invariant combinations of components $\eta_l$. The equilibrium values $\eta_l$ in the low-symmetry phase are determined by solving a system of state equations $\partial F / \partial \eta_l = 0$, where the index $l$ takes all values from 1 to m, here m is the dimension of the IR.

The study of inhomogeneous states in the Landau theory began with the Lifshitz paper [6], who investigated the stability of a homogeneous phase with the inhomogeneous fluctuations of the OP. In [6] it was assumed that: 1) a macroscopic crystal is characterized by low volume coordinate $\vec{X}$, and 2) the components of OP depend on the coordinates $\eta_l = \eta_l(\vec{X})$, and 3) nonequilibrium potential is a functional $F(\eta_l, \partial\eta_l / \partial X_j)$ of the OP components and its spatial derivatives, where the equilibrium distributions of the OP in the crystal correspond to extremals of the functional $F(\vec{X})$, and 4) phase transition at each point $\vec{X}$ goes though IR with the wave vector $\vec{k}_0$. In [6,7] it was proved that if a nonequilibrium Landau potential contains the Lifshitz invariant (antisymmetric quadratic combination, being linear in the relation to the OP components and its spatial derivatives) the highly symmetric phase becomes heterogeneous phase by the second-order transition. The temperature $T_0$ of the phase transition is above the Curie point $T_c$ of possible phase transition into the low-symmetric homogeneous phase.

The assumption of the locality of the OP transformation properties, namely dependence $\vec{k}_l = \vec{k}_l(\vec{X})$, was done in [4]. The assumption $\vec{k}_l = \vec{k}_l(\vec{X})$ is also acceptable, as well dependence of the OP on the coordinates $\eta_l = \eta_l(\vec{X})$. Indeed, in the classical Landau theory [5], the dependence of the OP on the coordinate is impossible, since the OP is equivalent to the harmonic coefficients of $\varphi_{\vec{k}_0 n_0 l}(\vec{x})$ in the expansion of the state density (1) of the IR basis functions (excluding the point symmetry, the expansion (1) is the Fourier integral). Consequently, the components of the OP $\eta_l$ and vector $\vec{k}_l$ are by definition independent of $\vec{x}$. Therefore, further heterogeneity of the functions will be written as a function of macro-coordinate $\vec{X}$, and not $\vec{x}$, to emphasize the macroscopic inhomogeneity. It is obvious that the dependence of OP on the coordinates in inhomogeneous models [1,2,6,7], using the calculus of variations, implied the existence of the local Landau potential, which depended on the macro-coordinate $\vec{X}$. In [8] has been formulated the principle of local homogeneity, which separates the coordinates and generalizes the Lifshitz formalism. The essence of the principle of local homogeneity in the Landau theory is the following: 1) locally in a macroscopic small volume characterized by macro-coordinate $\vec{X}$, the crystal can be regarded as homogeneous; 2) each macroscopic small volume with the coordinate $\vec{X}$ corresponds to an independent subspace $\{\vec{x}\}_{\vec{X}}$, where the Landau potential $F(\vec{X})$ can be constructed; 3) the variational theory for $F(\vec{X})$ is formulated in the macro space $\{\vec{X}\}$; 4) the inhomogeneous crystal state can be characterized as a relationship $\eta_l = \eta_l(\vec{X})$, or $\vec{k}_l = \vec{k}_l(\vec{X})$. The last provision differs the considered formalism from the Lifshitz approach. Moreover, if the components $\eta_l$ are variable parameters, the vector $\vec{k}_l$ describe the transformation properties of the OP under translations, and are not varied parameters. As shown in [4,8], the relationship $\vec{k}_l = \vec{k}_l(\vec{X})$ leads to the appearance of the additional variables in the Landau local theory as the tensor compensating fields, and not to the variation of $F$ in $\vec{k}_l$.

The most prominent examples of states with $\vec{k}_l = \vec{k}_l(\vec{X})$ are the states with the local translational symmetry. They can be obtained by deformation of the crystal lattice, which does not stretch and break with the formation of dislocations. These are the states studied by de Gennes description of the deformations in SmA [2]. The main variables varied, he used a comprehensive OP $\psi(\vec{X})$, $\tilde{\psi}(\vec{X})$ and the director $\vec{n} = \vec{n}(\vec{X})$ - the unit normal vector perpendicular to the layers of SmA. OP transformed as follows: $\hat{\vec{a}}\psi(\vec{X}) = e^{i\vec{k}\vec{a}}\psi(\vec{X})$, $\hat{\vec{a}}\tilde{\psi}(\vec{X}) = e^{-i\vec{k}\vec{a}}\tilde{\psi}(\vec{X})$. Since $\vec{k} = 2\pi\vec{n}/d$, the dependence $\vec{n} = \vec{n}(\vec{X})$ will automatically mean that the vector $\vec{k}$ is independent of $\vec{X}$. Here the $d$-spacing between layers in the SmA. In [2] it was assumed that $d = const$. Lo fact, de Gennes was the first who investigated the condition of a local transformation properties OP ($\vec{k} = \vec{k}(\vec{X})$) in the Landau theory.

The representations with $\vec{k} \neq 0$ where vector $\vec{k}$ characterizes the translational properties of the field function, have been poorly studied in the field theory (as in the field theory [3,9] are usually considered the finite-dimensional IR with $\vec{k} = 0$). At the same time, the study of non-trivial representation of the translation subgroup with $\vec{k} \neq 0$ plays a significant role in the crystallography and the physics of phase transitions. The compensating field appeared in the field theory in the extended derivative in connection with the construction of an invariant Lagrangian with respect to the abstract local gauge group of internal symmetries. In this paper the compensating fields appear in the extended derivative as a result of the requirement of the translational invariance of the Landau local potential. So the paragraph 7 will show that for the construction of the Ginzburg-Landau theory [1] the transformation properties of the components $\psi_\omega$ of the OP are set under the local transformation at translation time $\hat{\tau}$:

$$\hat{\tau}\psi_\omega = e^{i\omega(\vec{X})\tau}\psi_\omega. \tag{2}$$

Here, the parameter $\omega$ of the IR is the frequency, it characterizes the IR of the time translations subgroups. To construct the de Gennes model [2,8] (p.6) the OP transformation properties are given by the local transformation in the spatial translations $\hat{\vec{a}}$:

$$\hat{\vec{a}}\eta_l(\vec{X}) = e^{i\vec{k}_l(\vec{X})\vec{a}}\eta_l(\vec{X}). \tag{3}$$

Unlike the scalar electrodynamics model U(1) with the electromagnetic vector potential for the model with the vector group parameter of representation - $k_p(\vec{X})$, the compensating field - $A_{pi}$ is tensor, and the system of equations of state contains the equation of the continuum theory of dislocations [8,10,11]. The analogy of magnetostatics and the continuum theory of dislocations has long been known [11,12]. However, the disputes about the correspondence between the observed values are still going on. [13] What field corresponds to the electromagnetic potential $A_i$ - the potential of the stress field [11] or the distortion tensor [12]? In the first case the analogue of the current density $j_i$ is the density of dislocations $\rho_{pi}$ [11], in the second case - the stress tensor $\sigma_{pi}$ [12]. To select the physical interpretation of the compensating field for the local representation (3) is the main task of this article. In the p. 3 we prove that the compensating field of the OP with $\vec{k}_l = \vec{k}_l(\vec{X})$ is the distortion tensor. In the Landau local theory this proof is made possible due to the fact that the transformation properties of the compensating field are consistent with the transformation properties of the OP (3). Note that in [11-13], there was no OP or field function the derivative of which would have to be extended, so the analogue of the electromagnetic potential in the continuum theory of dislocations didn't act as the compensating field.

The modern gauge theory of dislocations also uses the distortion tensor [12.14] as the compensating field. However, its results are unconvincing, since it is linear and contains no interaction. The compensating field in [14] is a part of the Lagrangian term in combination with the spatial derivative of the displacement vector $u_p$: $D_j u_p = \partial u_p / \partial X_j + A_{pj}$, and does not appear in the extended derivative, as for example, in the models [8,9]:

$$D_j^l \eta_l = \left( \frac{\partial}{\partial X_j} - i \sum_p \kappa_p A_{pj}^l \right) \eta_l, \quad D_j^l \psi = \left( \frac{\partial}{\partial X_j} - i e A_j \right) \psi$$

(here $A_{pj}$ is the tensor distortion, $A_j$ is the electromagnetic potential). Therefore, the gauge theory of dislocations [12.14] has no the extension of the derivative as such. There is no interaction between the varied variables $u_p$ and $A_{pj}$ (there is no concept of the charge interactions like $\kappa_p$ and $e$.)

In the framework of the Landau local theory we have managed not only to reproduce the main results of the gauge theory of dislocations [12.14], to get the Newton's equations as equations of continuity and Peach-Koehler force [15], but also to trace the analogy with the Euler hydrodynamic equations p.4. In contrast to the gauge theory of dislocations [12.14], the Landau local theory, as well as in electrodynamics, there are solutions available for the compensating field - 4-distortion tensor ($A_{pj}, \upsilon_p$), here $v_i$ is the velocity. From the condition of Lorentz for the 4-distortion tensor in p. 4 we obtained the classical expression for the speed of sound, allowing to interpret ($A_{pj}, \upsilon_p$) as a phenomenological phonon potential. The paragraph p. 5 presents the derivation of equations of a particle motion in the phonon field out of the action for the phenomenological phonon potential, similar to the calculations of the equations of a particle motion in the electromagnetic field. It is demonstrated that the equations of a particle motion in the phonon field contain both the Peach-Koehler force and the Euler potential force of hydrodynamics.

As is known, the superconducting state in the external magnetic field is diverse and includes dislocations, see reviews [16,17] and the references provided. Therefore, there is reason to use the Landau local potential, which depends on the density of distribution of the electric charge with $\vec{k}_l = \vec{k}_l(\vec{X})$ for the description of superconductivity. The second objective of the presented paper is the investigation of the Landau local potential, the OP of which has the local transformation properties both (2) and (3). In this case, the OP extended derivative shall contain both the electromagnetic potential and the phenomenological phonon potential (distortion tensor). In this paper, we show that:

1) The OP interaction with the distortion tensor is a macroscopic way to describe the electron-phonon interaction, see p. 7.

2) The minimum magnetic flux in the Landau local theory for the states described by the OP with $\vec{k} \neq 0$ is equal to the quantum of magnetic flux, which is obtained in the standard description of a superconductor in the magnetic field [18]. But in the case of the OP with $\vec{k} \neq 0$, there is no need to use artificial doubling of the charge in the extended derivative [19], based on the microscopic BCS theory [20], see p. 8.

3) The low-symmetric solutions of the state equations, for which the inversion of space-time is not equivalent to the complex conjugation of the OP components, correspond to the d-symmetry of the superconducting gap in the reciprocal space. In this case, the superconducting gap is described by a quadratic invariant, which, for the crystals with a fourfold symmetry axis is represented as $I = 4\rho^2(1+\cos 4\theta)$. Here $I$ is the OP quadratic invariant, $\rho$ is the OP components module, and $\theta$ is the angle of turn around the axis, see p. 9.

## 2. Compensating Field

For models with $\vec{k} \neq 0$, an inhomogeneity of the sort $\vec{k}_l = \vec{k}_l(\vec{X})$ results in the nontrivial transformations of the OP spacial derivative $\partial \eta_l / \partial X_j$ under unit translations $\vec{a}$ in the subspace $\{\vec{x}\}_{\vec{X}}$ [7,8]. The case is:

$$\hat{\vec{a}} \frac{\partial \eta_l}{\partial X_j} = \frac{\partial}{\partial X_j} \left( e^{i\vec{k}^l(\vec{X})\vec{a}} \eta_l(\vec{X}) \right) = e^{i\vec{k}^l(\vec{X})\vec{a}} \left[ i \frac{\partial \vec{k}^l(\vec{X})\vec{a}}{\partial X_j} \eta_l(\vec{X}) + \frac{\partial \eta_l(\vec{X})}{\partial X_j} \right]. \quad (4)$$

As it is known [4], the inhomogeneous Landau potential is an invariant function constructed on the basis of OP and its spatial derivatives invariant with respect to the symmetry of the high-symmetry phase. It follows from (4) that the translation operator maps spatial derivatives of the OP, $\partial \eta_l / \partial X_j$ onto space of the OP itself $\eta_l(\vec{X})$ (first term in brackets of the expression (4)). By definition, components of the OP are the eigenfunctions of the translation operator (3). To construct invariants of the subgroup of translations which include spatial derivatives, we need first to construct the diagonal basis for the translation operator that includes the spatial derivatives of the OP. We apply the procedure of the derivative expansion which is proposed in the gauge field theory [3]. We extend spatial derivatives by introducing additional compensating field as a summand. We need to construct the extended derivatives so that they would be the eigenfunctions of the translation operator as well as OP components (3). According to (4) the compensating fields should be defined up to a gradient of the vector function. Let us show that this compensating field is a tensor.

The extended derivative is written down as:

$$D_j^l \eta_l = \left( \frac{\partial}{\partial X_j} - i \sum_p \kappa_p A_{pj}^l \right) \eta_l. \quad (5)$$

The translation properties of $\kappa_p A_{pj}^l$ shall be determined in such a way that extended derivative (5) should be the eigenfunction of the translation operator:

$$\hat{a}_q (\kappa_p A_{pj}^l) = \kappa_p A_{pj}^l + \delta_{pq} \frac{\partial k_p^l}{\partial X_j} a_q, \quad (6)$$

Here, $A_{pj}^l$ is the compensating field for component $\eta_l$ of OP, $\kappa_p$ is a phenomenological charge. From (6) results, that the transformation properties of the compensating field $A_{pj}^l$ are associated with the dimension of the vector $\vec{k}^l$. Since $A_{pj}^l$ should be transformed like $\partial k_p^l / \partial X_j$ with transformations from point symmetry group, and shall be the second-rank tensor. As is known, vectors in the star $\{\vec{k}\}$ of the irreducible representation (IR) – are dependent, since they are obtained from single vector $\vec{k}$ by operations of the point symmetry group of the high symmetry phase. Therefore, single independent tensor field $A_{pj}$, can be chosen such that it will compensate all vectors $\vec{k}^l$ of the IR star. For example [8], for the six-beam basis of the icosahedron

$$\vec{k}_1 = (0,1,h),\ \vec{k}_2 = (0,\bar{1},h),\ \vec{k}_3 = (h,0,1),\ \vec{k}_4 = (h,0,\bar{1}),\ \vec{k}_5 = (1,h,0),\ \vec{k}_6 = (\bar{1},h,0),$$

where $h = (\sqrt{5}+1)/2$, the extended derivatives take the form:

$$D_j^1 \eta_1 = \left[\frac{\partial}{\partial x_j} - i(\kappa_2 A_{2j} + h\kappa_3 A_{3j})\right]\eta_1,\quad D_j^2 \eta_2 = \left[\frac{\partial}{\partial x_j} - i(-\kappa_2 A_{2j} + h\kappa_3 A_{3j})\right]\eta_2,$$

$$D_j^3 \eta_3 = \left[\frac{\partial}{\partial x_j} - i(h\kappa_1 A_{1j} + \kappa_3 A_{3j})\right]\eta_3,\quad D_j^4 \eta_4 = \left[\frac{\partial}{\partial x_j} - i(h\kappa_1 A_{1j} - \kappa_3 A_{3j})\right]\eta_4,\quad (7)$$

$$D_j^5 \eta_5 = \left[\frac{\partial}{\partial x_j} - i(\kappa_1 A_{1j} + h\kappa_2 A_{2j})\right]\eta_5,\quad D_j^6 \eta_6 = \left[\frac{\partial}{\partial x_j} - i(-\kappa_1 A_{1j} + h\kappa_2 A_{2j})\right]\eta_6.$$

The extended derivatives with opposite vector $\vec{k}_l$ take the form: $\tilde{D}_j^l \tilde{\eta}_l$, here $\tilde{D}_j^l$ and $\tilde{\eta}_l$ means a complex conjugate. While constructing (7) we took into account that the change of coordinates of the vector $\vec{k}^l(\vec{X}) = \mu_i^l(\vec{X})\vec{e}_i$ is equivalent to the change of the values of basis vectors of reciprocal space $\vec{k}^l(\vec{X}) = \mu_i^l \vec{e}_i(\vec{X})$ in the subspace $\{\vec{x}\}_{\vec{X}}$. This allowed from the tensor star $\kappa_p A_{pj}^l$ in the extended derivative (5) to move to the vector star $\mu_p^l$ in the reciprocal space $(\mu_p^l \kappa_p) A_{pj}$ in (7). This example demonstrates the technique of constructing the extended derivatives for the IR space group.

Note that the $(\mu_p^l \kappa_p)$ is a scalar product of vectors, so the extended derivatives (7) incorporate the tensor components $(\mu_p^l \kappa_p) A_{pj}$ which, when converted from the point symmetry group, are transformed together with the components of the OP $\eta_l$. A similar situation holds for the extended derivative of the Ginzburg-Landau potential $D_j^l \psi = \left(\frac{\partial}{\partial X_j} - i\frac{2e}{\hbar c} A_j\right)\psi$ [18] in which the electromagnetic potential is converted together with the components of the wave function under inversion of time, while the derivative itself $\partial/\partial X_j$ is invariant under inversion of time. That is in the Landau local theory derivative the extended derivative is not a vector for the construction, in contrast to the gauge field theory (in the model U(1) the conversion of the compensating field under the inversion of time shall be defined from the second thought as the internal gauge symmetries of the Lagrangian function are not related with the symmetry of space-time [3,9]).

We can construct hence the eigenfunctions containing spatial derivatives of the OP in the form of extended derivatives for the translation operator. However, we introduce here an additional tensor compensating field that represents additional independent degrees of freedom. Besides, the symmetry-dependent interaction between compensating field and OP is determined by expressions (5, 7).

## 3. Observables. Stresses and Dislocations

States described by an OP with local translational properties $\vec{k} = \vec{k}(\vec{X})$, have the following simple interpretation: inhomogeneous deformation of crystal lattice with the local translational symmetry. Such deformation is generally accompanied by discontinuities and occurrences of dislocations of the lattice. Let us take the two-dimensional model $\vec{k}(\vec{X}) = (k_1(X_1, X_2), k_2(X_1, X_2))$, where $\vec{k} = (k_1, k_2(X_1))$ and the lattice period is connected with the vector $\vec{k}$: $\vec{a}\vec{k} = 2\pi$. If one combines two lattices with different periods we get discontinuities or incompatibilities of lattice, when the atom from one side will not match the atom on the other side. In a three-dimensional model, edge dislocations with lines perpendicular to the plane of reference are related to these incompatibilities. Here, the dislocations appear not under the random deformation of lattice but when two regions with different periods are "glued" that respects to representations about existence of local translational symmetry in the macroscopically small regions. For example, when the high-symmetric phase is invariant under all translations of the three-dimensional space from $\vec{k} = \vec{k}(\vec{X})$ results the following $\vec{a} = \vec{a}(\vec{X})$. Such a situation occurs in the transition of SmA nematic [2]. Therefore, the equations of state for the model with $\vec{k} = \vec{k}(\vec{X})$, must include the equations of the continuous dislocation theory [10,11]. Based on these concepts, we can define the observed values in our model.

Using equations (5,6) we are able to construct a translationally invariant inhomogeneous Landau potential as a function of the OP and its extended derivatives. In this model, the introduced compensating tensor field $A_{pj}$, is an independent variable, and a variation of the potential with respect to it must be equal to zero. Similar to electrodynamics, in the nonequilibrium potential it is necessary to consider the translational invariants of the compensating tensor field itself, which represent antisymmetric spatial derivatives of the second index (6):

$$e_{jkn}\left(\partial A_{pn} / \partial X_k\right). \tag{8}$$

where $e_{jkn}$ is third-rank antisymmetric tensor. Thus, the local Landau potential which is invariant with respect to elementary translation takes the form:

$$F(\vec{X}) = F(\eta_l \tilde{\eta}_l, D^l_j \eta_l \tilde{D}^l_j \tilde{\eta}_l, e_{jkn} \partial A_{pn} / \partial X_k). \tag{9}$$

Here an interaction between the OP and the compensating field is included into the extended derivatives.

There are two different models of the continuum theory of dislocations: Kroner's, see review [11], and Kadić-Edelen's [12,14]. Both theories trace the analogy of the continuum theory of dislocations with electrodynamics. The stress field potential $\Sigma_{pj}$ acts as the main variable in the first one, and the distortion tensor $w_{pj}$ - in the second one. We are to find out to what is equivalent to compensating field $A_{pj}$: $\Sigma_{pj}$ or $w_{pj}$.

Let's assume that the compensating tensor field is equivalent to the stress field potential: $A_{pj} \equiv \Sigma_{pj}$ introduced by Kroner to describe the continuous distribution of the stationary dislocations [11]. Since the equilibrium condition for the stationary states is given by $\partial \sigma_{pj}/\partial X_j = 0$, the stress tensor can be defined as the rotor:

$$\sigma_{pj} = e_{jkn}\left(\partial \Sigma_{pn}/\partial X_k\right). \tag{10}$$

In this case, the equations of state derived from the variation of the Landau local potential (9) on the components of the compensating field have the form:

$$\delta F/\delta A_{pj} = \partial F \bigg/ \partial A_{pj} - e_{jkn}\frac{\partial}{\partial X_j}\left(-\frac{\partial F}{\partial \sigma_{pn}}\right) = 0,$$ and coincide with the definition of the

dislocation density in the continuum theory of dislocations [10]:

$$\rho_{pj} = -e_{jkn}(\partial w_{pn}/\partial X_k). \tag{11}$$

Indeed, since $\partial F/\partial \sigma_{pj} = w_{pj}$ is the distortion tensor (here $w_{pj}$ is a generalization of the deformation tensor for the states with dislocations), the equation of state $\delta F/\delta A_{pj} = 0$ results in $\partial F/\partial A_{pj} = \rho_{pj}$ being the density of dislocations (11). In this case, the Burgers vector is defined as the integral $B_i = \int_{S_L} \rho_{ij} ds_j$ over the surface bounded by the contour $L$. Kroner in his theory saw an analogy between: the current density $j_i$ and density of dislocations $\rho_{pi}$, electromagnetic potential $A_i$ and the potential of the stress field $\Sigma_{pi}$, magnetic induction $B_i$ and stress tensor $\sigma_{pi}$ [11].

Let's draw our attention to the fact that in the present model for the potential (9), there are two vortex expressions (10) and (11) which are associated with the observables (invariant values with respect to the elementary translation), one of which is the definition of (8), and the second one is the equation of state $\delta F/\delta A_{pj} = 0$. Therefore, generally speaking, there are two possible interpretations of the compensating field. [21]

The alternative definition of the compensating field $A_{pj} \equiv w_{pj}$ as a distortion tensor was introduced by Kadić and Edelen [12]. In this case: 1) the determination of the dislocation density (11) coincides with the definition of the observed translational invariants of the compensating field (8), and 2) the expression (10), which is the condition of equilibrium in the steady state is determined from the equations of state, and 3) the conjugate values are defined as follows: $\partial F/\partial A_{pj} = \sigma_{pj}$ and $\partial F/\partial \rho_{pj} = -\Sigma_{pj}$.

The following methodological considerations are in favor of the definition $A_{pj} \equiv w_{pj}$.
1) The elastic distortion tensor in the equation (10), as well as a compensating field tensor (6), is defined up to a gradient of the vector function $\partial u_p/\partial X_j$ when determining dislocations density (11).
2) The expression (11) is the definition of the dislocation density [10], and the expression (10) is a consequence of the equilibrium in the steady state. Condition (10) is determined from the equations of state, and, generally speaking, is not a definition of the stress tensor (in dynamics $\partial \sigma_{pj}/\partial X_j \neq 0$).
3) The elastic distortion tensor $w_{pj}$ is a tensor of the second rank and not a pseudotensor of the second rank as the stress field potential $\Sigma_{pj}$ in (10). Therefore, in determining $A_{pj} \equiv w_{pj}$ there is no need to enter the pseudoscalar phenomenological charge [21] in order to conform with the mathematical dimension of the compensating field in the extended derivative (5,6,7).

4) The above illustration of the two-dimensional lattice clearly shows that the lines of the edge dislocations are perpendicular to the change plane of the group parameter $\vec{k}(\vec{X})$. It follows that the distortion $w_{pj}$ is a compensating field $A_{pj}$.

Indeed, let's suppose that $k_j = (0, k_2(X_1, X_2), 0)$, and recognize that for the two-dimensional model $\partial w_{21}/\partial X_3 = \partial w_{22}/\partial X_3 = 0$, $w_{23} = 0$. Then the dislocation density is represented as: $\rho_{21} = -(\partial w_{23}/\partial X_2 - \partial w_{22}/\partial X_3)$, $\rho_{22} = -(\partial w_{21}/\partial X_3 - \partial w_{23}/\partial X_1)$, $\rho_{23} = -(\partial w_{22}/\partial X_1 - \partial w_{21}/\partial X_2)$. So $\rho_{2j} = (0, 0, \rho_{23})$. On the other hand, by definition (6), the compensating field for $k_2(X_1, X_2)$ is represented as $A_{2j} = (A_{21}, A_{22}, 0)$. Consequently, the tensor compensating field $A_{2j}$ and density of dislocations $\rho_{2j}$ can not be conjugate. Where from $\partial F/\partial A_{pj} \neq -\rho_{pj}$, and $A_{pj} \neq \Sigma_{pj}$. Hence, only variant of the above discussed interpretation of the compensating field is consistent when $A_{pj} \equiv w_{pj}$ and $A_{pj}$ is associated with the stress field $\partial F/\partial A_{pj} = \sigma_{pj}$.

The definition of the interpretation of the compensating tensor field entered in the Landau local theory in p. 2 is the key issue in the theory. Why, despite the obvious argument 1) -4) we have long adhered to the Kroner's model [11], and determined the compensating field as the potential of the stress field (10)? First, because of the Kroner's analogy with the magnetostatics where the current density $j_i$ corresponds to the dislocation density $\rho_{pi}$, and the magnetic induction $B_i$ - to the stress tensor $\sigma_{pi}$, which seems logical. The Kadić-Edelen's model [12] on the contrary: the current density $j_i$ is associated with the stress tensor $\sigma_{pi}$ and the magnetic induction $B_i$ with the density of dislocations $\rho_{pi}$. Against the Kadić-Edelen interpretation are the authors reviewed in [13], who believe that dislocations can not be associated with the massless fields of interaction, corresponding to the magnetic fields. The pseudoscalar charge dislocation [21] resulted from these arguments in the Landau local theory. In contrast to [11,13], where the OP was not used, in [21] it was assumed that dislocations can't be present without the OP. In addition, in [21] it was necessary to give a reasonable interpretation of the solutions of the free field $A_{pj}$ similar to the solutions of the Maxwell equations for the free electromagnetic field. The analogy with the theory of elasticity where the main variable was the stress tensor in the Landau potential seems logical. [8] The analogy with the theory, where the Landau potential (9) was the functional density of dislocations (11), was not possible because there was no reasonable interpretation of the dislocation density, for example, in the air. This was perhaps the most weighty argument similar to the argument of the authors in [13] against the use of the distortion tensor as the compensating field.

At the same time, there was a hypothesis about the description of sound through a 4-distortion tensor ($A_{pj}, \upsilon_p$), where $\upsilon_p$ is the component of the velocity field [12], as a pair of strength fields: $(\rho_{pj}, \varepsilon_{pj}) \equiv (-e_{jkn}(\partial A_{pn}/\partial X_k), (-\partial \upsilon_p/\partial X_j + \partial A_{pj}/\partial T))$, by analogy with the electromagnetic wave intensity $(B_j, E_j) \equiv (e_{jkn}(\partial A_n/\partial X_k), (-\partial \varphi/\partial X_j - \partial A_j/\partial T))$ (derivation of the expressions of phonon field intensity, see p. 4, 5). The hypothesis was based on heuristic ideas about the sound as the gradient of the velocity field $-\partial \upsilon_p/\partial X_j$ which is easily observed, for example, when playing a musical instrument. Indeed, strings and membranes in musical instruments are mounted on the edges, so the vibrations there of are accompanied by a nonuniform distribution of the air velocity. And all the wind musical instruments have expansion (or contraction) on the end, so the air is blown out of them under the law $-\partial \upsilon_p / \partial X_j \neq 0$. As is known, the formation of dislocations is also accompanied by sound. However, the existence of dislocations in the air, according to the definitions [10], was doubtful. The situation changed when the additional Lorentz conditions (see (19) and [14]) required for the wave solutions of equations of state were applied to the compensating field of the 4-distortion tensor (The same situation occurs in electrodynamics in obtaining wave solutions for electromagnetic fields). Since the Lorentz conditions have the form of the equations of continuity, the distortion tensor $A_{pj}$ is equivalent to the flux of the velocity field $\upsilon_p v_j$. Representation of the sound as the vortex flux of the velocity field $e_{ikj}(\partial(\upsilon_p v_j)/\partial X_k)$ accepts the interpretation of the 4-distortion tensor as a phenomenological phonon potential.

The description of the superconductivity phenomenon is also in favor of the determination of the distortion tensor as a phenomenological phonon potential. The most well-known theory which deals with the interaction with phonons is the BCS theory [20]. It is known [18.20], the description of superconductivity takes into account both the electromagnetic interaction and the electron-phonon interaction. You can hold the following analogy. First, the received minimal OP interaction and the tensor compensating field (5.7) in the form is similar to the electromagnetic interaction and describes the elastic properties of the condensate through the distortion tensor. Secondly, the invariance of Landau local potential with respect to elementary translations of both spatial and temporal (2,3) results that the extended derivative must contain both the phonon field $A_{pj}$ in the form of the distortion tensor and the electromagnetic field $A_j$ (see paragraph .7).

So, despite the fact that in determining the dislocation we treated the states with the local translation symmetry, described by the OP with $\vec{k} = \vec{k}(\vec{X})$, the very definition of dislocations, as is known in [10], is formulated in the framework of the continuum medium. The determination of dislocations is not related to the definition of the local OP. The above heuristic arguments show that the distortion can be considered as a phenomenological phonon potential, similar to the electromagnetic potential. Moreover, the analogy between the dislocation density $\rho_{pi}$ and the magnetic induction $B_i$, and distortion $A_{pj} \equiv w_{pj}$ and electromagnetic potential $A_j$, as well as between the stress tensor $\sigma_{pi}$ and the current density $j_i$, which was not possible [13.21], proved to be the only consistent. We show that this analogy is evident when considering the dynamic case.

In the dynamic case, the equilibrium condition $\partial \sigma_{ij}/\partial X_j = 0$ for the statics should be transformed into the Newton's equations for dynamics:

$$\partial \sigma_{ij}/\partial X_j = \partial p_i/\partial T, \tag{12}$$

here $p_i$ is the momentum, $T$ is the macroscopic time. Therefore, the expression (10) can not serve as a definition of the stress tensor, since $\partial \sigma_{ij}/\partial X_j \neq 0$, the definition must be extended, in contrast to the dislocation density (11). In addition, in the dynamic model the stress tensor must match its original definition $\partial \sigma_{ij}/\partial X_j = f_i$ - as a tensor the divergence of which is force. [10]. Since the stress tensor is included explicitly in the equation of state, the equation (12), which is the differential recording of the law of conservation of momentum, must be a consequence of the system of equations of state. We show that (12) is the continuity equation, induced by the invariance of the Landau local potential with respect to the subgroup of elementary translations, in determining the stress tensor as: $\sigma_{pj} = \partial F/\partial A_{pj}$. Indeed, assuming that the process of change in physical quantities is slow enough so that the equilibrium state in a macroscopic time $T$ are achievable (the principle of local uniformity for time), let's take into account the time derivatives of the OP in the phenomenological potential. By analogy with the field theory of electrodynamics, we extend the time derivatives of the OP by introducing into the theory additional vector compensating fields $\upsilon_p^l$:

$$D_0^l \eta_l = \left( \frac{\partial}{\partial T} - i \sum_p \kappa_p \upsilon_p^l \right) \eta_l, \qquad (13)$$

$$\hat{a}_q (\kappa_p \upsilon_p^l) = \kappa_p \upsilon_p^l + \delta_{qp} \frac{\partial k_p^l}{\partial T} a_q \qquad (14)$$

(the plus sign on the right side (14) is consistent with the used below gauge condition (19) and follows from the equations of a particle motion in the phonon field (see p.5).) Transformations (6,14) correspond to the gauge transformation of the 4-vector of the electromagnetic potential ($A_j, \varphi$) in electrodynamics [9]:

$$g(eA_j) = eA_j + \partial\alpha / \partial X_j, \quad g(e\varphi) = e\varphi - \frac{1}{c} \partial\alpha / \partial T, \qquad (15)$$

where $e$ is the electron charge (here it is assumed that $\hbar=1$, $c=1$ [9]). In (15) $\alpha$ is a scalar function characterizing the gauge transformation of the field: $\psi: g(\psi) = e^{i\alpha}\psi$. As is the case with the tensor compensating field $A_{pj}$, we can omit the index $l$ in $\upsilon_p$, given that the time derivatives of all th OP components can be compensated by the introduction of a vector field $\upsilon_p$ for the IR star $\{\vec{k}\}$. The coefficients before $\upsilon_p$ of the extended derivative can be obtained in the same manner as in (7), in each case. Then the translation-invariant potential is a functional:

$$F(\vec{X}) = F(\eta_l \tilde{\eta}_l, D_j^l \eta_l \tilde{D}_j^l \tilde{\eta}_l, D_0^l \eta_l \tilde{D}_0^l \tilde{\eta}_l, \rho_{pj}, \varepsilon_{pj}). \qquad (16)$$

Here $\rho_{pj} = -e_{jkn} \partial A_{pn}/\partial X_k$ is the dislocation density (11), and

$$\varepsilon_{pj} = -\partial \upsilon_p / \partial X_j + \partial A_{pj} / \partial T \qquad (17)$$

Is the translational invariant (6.14), similar to the electric field intensity, which we associate with the phonon field intensity. The structure of non-equilibrium potential (16) is similar to the expression for the Lagrangian field theory of electrodynamics. Therefore, from the variation $F(\vec{X})$ of the compensating fields $A_{pj}$, $\upsilon_p$, we get the equations of state, similar to the Maxwell equations. Determination of the stress tensor $\partial F / \partial A_{pj} = \sigma_{pj}$ is consistent with the definition of the current density $\partial F / \partial A_j = j_j$ in electrodynamics. From the system of equations of state $\delta F / \delta A_{pj} = 0$, $\delta F / \delta \upsilon_p = 0$, using (11,16,17), we obtain the continuity equation for the stress tensor, repeating the operations related to the derivation of the continuity equation for the current. We introduce the designation $\partial F / \partial \upsilon_p = p_p$ and act on the equation $\delta F / \delta A_{pj} = 0$ by the divergence operator over the second index, and the equation $\delta F / \delta \upsilon_p = 0$ we differentiate with respect to time. Then from the first equation we get

$\partial \sigma_{pj} / \partial X_j + \partial^2 (\partial F / \partial \varepsilon_{pj}) / \partial T \partial X_j = 0$, and from the second

$\partial p_p / \partial T + \partial^2 (\partial F / \partial \varepsilon_{pj}) / \partial X_j \partial T = 0$ (in the first equation is taken into account that the divergence of the rotor is equal to zero $e_{jki} \partial^2 (\partial F / \partial \rho_{pi}) / \partial X_k \partial X_j = 0$). Subtracting the second from the first expression, we obtain (12). It is obvious that $p_i$ is the momentum which implies that $\upsilon_i$ is the field of the velocity. Thus, the definition $A_{pj} \equiv w_{pj}$ leads to the continuity equation (12), which is a differential recording of the law of conservation of momentum. The continuity equations (12) are related to the invariance of the Landau non-equilibrium potential with respect to the subgroup of elementary translations, which is similar to the law of conservation of momentum in classical mechanics.

For more obviousness we reviewed above an example of the state with the local translational symmetry. When the translational symmetry of the high-symmetric phase is discrete and not continuous, the low-symmetric state with $\vec{k} = \vec{k}(\vec{X})$, in general, does not have a local translational symmetry. This is because the vector $\vec{k}(\vec{X})$ does not correspond to the selected points of the Brillouin zone in each local subspace $\vec{X}$ as a result of the continuous dependence $\vec{k} = \vec{k}(\vec{X})$. However, all the above mentioned arguments are valid for such conditions since the IR is characterized by the vector $\vec{k}$ both for continuous and discrete subgroup of translations. An example of localization of the translation subgroup of the high-symmetric phase is implemented in the deformed state SmA [2] in the nematic-SmA transition. The transition of a crystal to the superconducting state in the external magnetic field is an example of violation of the translational symmetry of the discrete subgroup of translations.

## 4. The analogy with electrodynamics.

Let's draw to analogy between the Landau local theory, in which as a compensating field stands 4-distortion tensor: ($A_{pj}, \upsilon_p$) and electrodynamics. It is known [3,9] that the wave solutions of the Maxwell's equations for the free electromagnetic field appear in case of the Lorentz additional gauge condition:

$$\frac{\partial A_j}{X_j} = -\frac{1}{c}\frac{\partial \varphi}{\partial T}, \qquad (18)$$

where ($A_j, \varphi$) is the electrodynamic 4-vector, and $c$ is the speed of light. To get the wave solutions of state for the potential (16) in the absence of the OP $F = F_f(\rho_{pj}, \varepsilon_{pj})$, we introduce additional conditions for the compensating fields similar to (18):

$$\frac{\partial A_{pj}}{X_j} = \frac{1}{c_w^2}\frac{\partial \upsilon_p}{\partial T}, \qquad (19)$$

where $c_w$ is the wave velocity (we do not consider the anisotropy for simplicity). Square of the speed of sound $c_w$ in (19) follows from the expression dimension, and is linked to the lack of normalization $1/c_w$ in the transformation (6,14), in contrast to (15).

It is easy to see that the velocity $c_w$ is a function of the phenomenological coefficients of the potential $F_f = F_f(\rho_{pj}, \varepsilon_{pj})$. Indeed, $c_w^2 = \beta_1/\beta_2$, where $\beta_1$ is the coefficient of the quadratic invariant composed of the spatial derivatives $\rho_{pj} = -e_{jkn}(\partial A_{pn}/\partial X_k)$, and $\beta_2$, is the coefficient of the quadratic invariant composed of: $\varepsilon_{pj} = -\partial \upsilon_p/\partial X_i + \partial A_{pi}/\partial T$. Phenomenological coefficients $\beta_1$ and $\beta_2$ have the physical dimension of pressure and density, respectively. As in the field theory, the condition (19) results in the d'Alembert equation for group parameter $k_p$:

$$\Delta k_p - \frac{1}{c_w^2}\frac{\partial^2 k_p}{\partial T^2} = 0 \tag{20}$$

It results, if we act on (19) with the operator of the elementary translation. The equation (20) means that the wave solutions in the crystal for a free 4-distortion tensor exist in a state in which the components of the vector $k_p$ change under the wave law with the velocity $c_w$ (20). The converse is also true: if the vector $\vec{k}$ satisfies the equation (20), the components of the field ($A_{pj}, \upsilon_p$) are related by the additional condition (19). At the coincidence of the phenomenological parameter $c_w = \sqrt{\beta_1/\beta_2}$ and the parameter in (20) the components of the field ($A_{pj}, \upsilon_p$) satisfies the d'Alembert wave equation with the operator $\Delta - \frac{1}{c_w^2}\frac{\partial^2}{\partial T^2}$. The change of the vector $k_p(\vec{X},T)$ under the law (20), for a model with the local translation symmetry can be interpreted as a change of the lattice spacing $a_p(\vec{X},T)$ under the wave law. This suggests that ($A_{pj}, \upsilon_p$) is a phenomenological phonon field. The wave equations for the free phonon field are consistent with the mechanical wave vibrations (acoustic, since $c_w = const$). But the phonon field itself is not the mechanical vibrations of the medium, which are usually described as a change in the density under the wave law [22]. If we follow the analogy with the electromagnetic wave ($B_j, E_j$), the sound will be a pair ($\rho_{pj}, \varepsilon_{pj}$) of the dislocation density and intensity of the phonon field, each component of which satisfies the wave equation under the condition (19).

We write for the three-parameter model of the Landau local theory (with a vector parameter $\vec{k} = \vec{k}(\vec{X})$) the expressions similar to the Coulomb force and the Lorentz force for a one-parameter gauge model of the electrodynamics (with the scalar parameter $\alpha = \alpha(\vec{X})$ (15)). This will show how the intensity of the phonon field ($\rho_{pj}, \varepsilon_{pj}$) is implemented.

Since the law of conservation of the electric charge corresponds to the law of conservation of the momentum, the analog of the Coulomb potential force is the expression:

$$f_j = p_i \varepsilon_{ij} = -p_i \partial \upsilon_i / \partial X_j . \tag{21}$$

We show that (21) is a potential force in the Euler equations of hydrodynamics. By substituting (21) in the usual form $-p_i \partial v_i/\partial X_j$ (where $\vec{v} \equiv \vec{\upsilon}$) in the equations of motion and given the pressure $p$, we obtain the Euler equations for the potential fluid flux [22]:

$$\rho^{den}\partial v_j / \partial T = -p_i \partial v_i / \partial X_j - \partial p / \partial X_j . \tag{22}$$

Where, the Bernoulli equation results in the steady state:

$$v^2/2 + p/\rho^{den} = const , \tag{23}$$

were $\rho^{den}$ - density.

Also let's find an analogue of the Lorentz force $f_i = e_{ijk} j_j B_k$, where $B_k = e_{klm} \partial A_m / \partial X_l$. We substitute in the expression for the Lorentz force the current density $j_j$ for the stress tensor $\sigma_{nj}$, and for the magnetic induction $B_k$ the density of dislocations $\rho_{nk}$, and we obtain:

$$f_i = e_{ijk}\sigma_{nj}\rho_{nk}, \tag{24}$$

here $\rho_{nk} = -e_{klm}\partial A_{nm}/\partial X_l$ (11). The expression (24) is the force acting on the dislocation in the case of a continuous distribution of the density thereof. The force (24) is the Peach-Koehler force [10,13], represented as:

$$f_i = e_{ikj}\tau_k b_n \sigma_{nj}^{ext}. \tag{25}$$

Indeed, given that the tensor $e_{ikj}$ is antisymmetric, the permutation of indices will change the sign (25), reflecting the transition from the external stress $\sigma_{nj}^{ext}$ in (25) to the internal stresses in $\sigma_{nj}$ (24). The density of dislocations $\rho_{nk}$ is the generalization of the expression $b_n\tau_k$ for a continuous distribution of dislocations [10].

Thus, by analogy with electrodynamics we obtained two known forces (21,24), at first glance, from different areas of physics. One would get the vortex term in the Euler hydrodynamics equations on the analogy with the Lorentz force. We show that the force $\vec{f} = \rho^{den}[\vec{\upsilon}\times(\vec{\nabla}\times\vec{\upsilon})]$ describing the vortex fluid flux in the Euler equations is contained in the expression (24) subject to (19). Indeed, from the Lorentz condition (19), which has the form of the equations of continuity, it follows that the distortion tensor $A_{nm}$ can be represented as the flux of the velocity field $c_w^{-2}\upsilon_n v_m$, where $v_m$ is the flux velocity. Then the density of dislocations will be represented as: $\rho_{nk} = -e_{klm}c_w^{-2}\partial(\upsilon_n v_m)/\partial X_l$. Given that in the fluid $\sigma_{ij} = -\delta_{ij}p$, and the flux velocity $v_m$ has the same direction as the velocity $\vec{v} = \chi\vec{\upsilon}$, out of (24) we obtain $\vec{f} = \chi p c_w^{-2}[\vec{\upsilon}\times(\vec{\nabla}\times\vec{\upsilon})]$. The resulting force expression is equivalent to the vortex term in the Euler equations provided when:

$$c_w^2 = \chi p/\rho^{den}. \tag{26}$$

This is a well-known Newton's equation for the speed of sound. From (26) follows that the dimensionless coefficient $\chi$ corresponds to a correction in the Newton's equation associated with the ratio of heat capacities [22]. Note that the vortex term of the Euler equation is a summand in the expression for the Peach-Koehler force subject to (19.26). The expression (24) is exactly equivalent to the vortex term in the Euler equations if $v_j\partial\upsilon_j/\partial X_i - v_i\partial\upsilon_j/\partial X_j = 0$.

The classical expression for the speed of sound can be obtained directly from the Hooke's law and the condition (19). Indeed, since the stress tensor is proportional to the elastic distortion tensor, as a generalization of the deformation tensor, then $\sigma_{ij} = -KA_{ij}$, here $K$ is the elastic modulus. Then, from the continuity equation (12) $\sigma_{ij} = -p_i v_j = -\rho^{den}\upsilon_i v_j$, and from the Lorentz condition (19): $A_{ij} = c_w^{-2}\upsilon_i v_j$, it follows that: $c_w^2 = K/\rho^{den}$, whence $c_w = \sqrt{K/\rho^{den}}$.

5. **Particle motion of the phonon field**

The above calculations suggest that the expression for the phonon field intensity can be derived from the principle of least action. As is known, the action for the charge in the electromagnetic field is represented as [23]:

$$S_e = -\frac{e}{c}\int_a^b (-A_j dx^j + \varphi\, cdt). \tag{27}$$

From the expression (27) and the principle of least action we obtain the expressions for the Coulomb and Lorentz forces. By analogy, we assume that the action for the phenomenological charge in the phonon field $A_{ij}$ is:

$$S_{ph} = -\hbar\kappa_i \int_a^b (A_{ij} dx^j + \upsilon_i dt), \tag{28}$$

where $\kappa_i$ is the phenomenological vector charge of the particle (5), which has the dimension [1/m], and the multiplier $\hbar$ (Planck's constant) is introduced from the dimensional considerations and for the convenience of further calculations. In (28) the is no multiplier $c_w$ in front of the differential $dt$, as $\upsilon_i dt$ is independent of the speed of sound $c_w$ and has a dimension of [m]. This is due to the conjugation between the momentum $p_i$ and speed $\upsilon_i$ in theory. Then, integrating with respect to time under the integral (28), and given that $dx^j/dt = v^j$, we supplement the Lagrangian function conditioned by the phonon potential:

$$L_{ph} = -\hbar\kappa_i A_{ij} v^j - \hbar\kappa_i \upsilon_i. \tag{29}$$

Consequently, the generalized momentum $dL/dv^j$ [23], taking into account the phonon contribution, will be:

$$P_j = p_j - \hbar\kappa_i A_{ij}, \tag{30}$$

If you take into account the electromagnetic potential in $dL/dv^j$, the generalized momentum will be represented as:

$$P_j = p_j + \frac{e}{c}A_j - \hbar\kappa_i A_{ij}. \tag{31}$$

From the Euler-Lagrange equations $\dfrac{d}{dT}\dfrac{\partial L}{\partial v_j} = \dfrac{\partial L}{\partial X_j}$ for a particle in the phonon field (29.30) we obtain:

$$\frac{dp_j}{dT} - \hbar\kappa_i \frac{dA_{ij}}{dT} = -\hbar\kappa_i \frac{\partial}{\partial X_j}(A_{ij} v^j) - \hbar\kappa_i \frac{\partial \upsilon_i}{\partial X_i}. \tag{32}$$

Whence, differentiating with respect to parts the second term on the left side of (32): $\dfrac{dA_{ij}}{dT} = \dfrac{\partial A_{ij}}{\partial T} + \dfrac{\partial A_{ij}}{\partial X_k}v^k$ and transforming the first term on the right side under the well-known formula of the differential geometry (given that $v^j$ are $X_j$ are the independent variables):

$$\frac{\partial}{\partial X_j}(A_{ij} v^j) = \frac{\partial A_{ij}}{\partial X_j}v^j + e_{jmn}e_{npq}v_m \frac{\partial A_{iq}}{\partial X_p}, \text{ we obtain:}$$

$$\frac{dp_j}{dT} = \hbar\kappa_i \frac{\partial A_{ij}}{\partial T} - \hbar\kappa_i \frac{\partial \upsilon_i}{\partial X_j} + \hbar\upsilon_i e_{jmn}e_{npq}v_m \frac{\partial A_{iq}}{\partial X_p}. \tag{33}$$

Since the dimension of the phenomenological charge $\kappa_i$ is the same as that of the wave vector (the distortion tensor $A_{ij}$ is dimensionless) then $\hbar \kappa_i = p_i$ is the particle wave momentum. It performs as the charge in the phonon field. Indeed, if we assume that $\hbar \kappa_i = p_i$ then the first two terms on the right side of (33) is the force:

$$p_i \varepsilon_{ij} = p_i \frac{\partial A_{ij}}{\partial T} - p_i \frac{\partial \upsilon_i}{\partial X_j} \,. \tag{34}$$

It follows that the expression (17) is the intensity for the momentum $\hbar \kappa_i = p_i$. As shown above the (34), in the steady state, results in the Bernoulli equation (23). The third term in (33) is the Peach-Koehler force (24), since $e_{jmn} p_i v_m e_{npq} \frac{\partial A_{iq}}{\partial X_p} = e_{jmn}(-\sigma_{im})(-\rho_{in}) = e_{jmn}\sigma_{im}\rho_{in}$. This implies that the dislocation density is the intensity for the flux momentum $p_i v_m = -\sigma_{im}$ (12), or the stress tensor.

Thus, the definition $\hbar \kappa_i = p_i$ leads to a force (34) similar to the centrally symmetric force, and the Peach-Koehler force (24) similiar to the Lorentz force.

This conclusion confirms the correctness of the analogy with electrodynamics (p.4), and the correct choice of the transformation properties (6,14), which lead to intensity (17) and the Lorentz gauge (19). The last statement is fundamentally differs the Landau local theory from the Kadić-Edelen gauge theory. As in [12] the transformation properties of the distortion tensor $A_{ij}$ and the velocity vector $\upsilon_i$ lead to pseudo-Lorentz gauge conditions [14], see paragraph 10.

Momentum as the charge appeared in the Landau local theory due to the invariance of the non-equilibrium potential with respect to elementary translations, given in the local subspaces $\{\vec{x}\}_{\vec{X}}$. From this conclusion, it follows that this momentum can be associated with the quantum-mechanical wave momentum: $p_i = \hbar \kappa_i$. In fact, the OP with $\omega \neq 0$ and $\vec{k} \neq 0$ represents the symmetry of the wave function, as it is under the action of the operator of the time translation and space its wave properties manifest. The assumption of the existence in each macroscopic small area $\{\vec{x}\}$ with the coordinate $\vec{X}$ of the translational symmetry with the local period $a_p$, in our view, is equivalent to the quantum mechanical postulate that at short distances all particles have wave properties. In quantum mechanics this resulted in the replacement of momentum and energy by the differential operators in the equations of motion. In the Landau local theory the local transformation properties of the OP (3) lead to the minimal interaction, namely the extension of the derivatives $\partial \eta_l / \partial X_j$, $\partial \eta_l / T$, and the introduction to the theory of additional compensating fields ($A_{ij}, \upsilon_i$). The definition of the phenomenological charge interaction of the OP and the phonon field in the form of $\kappa_i = p_i / \hbar$ evidences the relationship of the two approaches. In paragraph 7 it will be shown that the electromagnetic interaction is a consequence of the locality of the transformation properties of the OP in relation to the time translations (2).

## 6. Minimal interaction in the Landau theory. De Gennes model. .

The most famous model in the Landau phenomenological theory, which uses the extended derivative, and accounts the dislocations in a deformed state is the de Gennes' model [2]. De

Gennes tried to construct the phenomenological potential for SmA, similar to the Ginzburg-Landau potential [1] to describe the screening effect of the stress field by the elastic dislocations similar to Meissner effect [18]. The phenomenological potential in [2] was represented as:

$$F = a|\psi|^2 + \frac{b}{2}|\psi|^4 + |(\vec{\nabla} - ik_0 \vec{\delta n})\psi|^2 + \Phi(\vec{n}),\qquad(35)$$

where $\Phi(\vec{n}) = c_1(\vec{\nabla}\vec{n})^2 + c_2(\vec{n}(\nabla \times \vec{n}))^2 + c_3(\vec{n} \times (\vec{\nabla} \times \vec{n}))^2$ is the Franck potential, chosen as the elastic potential. De Gennes considered a model in which the distance between the layers in the SmA won't change ($d = const$), from which it followed that in the deformed state $\vec{n} = \vec{n}(\vec{X})$. However, he chose the vector $\vec{\delta n}$ as a compensating field, and not the tensor, since he didn't check the translation invariance of the constructed potential. It is easy to see that the de Gennes potential is not invariant with respect to the elementary translation operator, since the vector field $\vec{\delta n}$ can not compensate changes of the director $\vec{n}(\vec{X})$ in three directions (4,5,6). Moreover, the variation $\vec{\delta n}$ is the observed value, it is invariant under elementary translations, and can not act as the compensating field. The authors [24] pointed out the observability of the compensating field in [2] and tried to introduce the abstract vector compensating field into the de Gennes theory, similar to the electromagnetic potential. It is known [2,24], the problem of screening of the stress field in the de Gennes model could not be solved for another reason, that is the usage of the Franck potential $\Phi = \Phi(\vec{n}(\vec{X}))$ as the elastic potential, because it contains irrotational terms of the director divergence type (35), which does not permit to obtain the equations similar to the London equations [18]. The usage of the nematic Franck potential to describe the elastic properties of the SmA, generally speaking, it is hardly possible, since it does not reflect the periodic structure of the SmA [10]. De Gennes introduced it for the phenomenological description of the deformed SmA, most likely because the director change leads to dislocations. Indeed, if we take the closed circuit $L$, the integral: $q_L = \oint_L \frac{\vec{n}}{d} d\vec{r}$ specifies the number of dislocations $q_L$ bounded by this circuit, here $d = 2\pi/k_0$ is the distance between the layers, $\vec{k}_0$ is the wave vector of the undeformed SmA [2].

It was demonstrated above that the theory with the compensating field - the distortion tensor, describes the state with dislocations and contains only vortex invariants of the compensating field in the Landau potential (9). Therefore, the model with the tensor compensating field [8,25] is free from the de Gennes drawbacks.

As in SmA the components $\sigma_{3j}$ are nonzero [10], the formalism implies that [8]:

$$\frac{\partial F}{\partial A_{3j}} = \sigma_{3j} = -i\beta_3 \kappa_3 \left(\psi_{-\vec{k}} \frac{\partial \psi_{\vec{k}}}{\partial X_j} - \psi_{\vec{k}} \frac{\partial \psi_{-\vec{k}}}{\partial X_j}\right) - 2\beta_3 \kappa_3^2 \psi_{\vec{k}} \psi_{-\vec{k}} A_{3j},\qquad(36)$$

here $\beta_3$ is the coefficient in front of the extended derivative in the potential. In approximation $|\psi_{\vec{k}}| = const$, by acting on (36) with the rotor operator over the second index, we obtain equation similar to the London equations if electrodynamics [18], which implies that the solutions of the equations of state $\delta F/\delta A_{3j} = 0$ describe the effect of shielding of the dislocation field with the internal stress:

$$\sigma_{3j} = \sigma_{3j}^0 \exp(-X_3/\delta),\ \rho_{3j} = \rho_{3j}^0 \exp(-X_3/\delta),\qquad(37)$$

where $\delta$ is the depth of penetration of the field perpendicular to the layers, and $\sigma^0_{3j} = const$, $\rho^0_{3j} = const$. We draw attention to the fact that the solutions for the observed values (37) have the same form of the exponential dependence. So it does not matter whether the internal stress shields the dislocation or dislocations screen the internal stresses, they decay together (37).

In describing the screening in SmA it is not necessary to adhere to $d = const$ of the de Gennes model, as the Franck elastic energy is not used here. In [8], we review the case when $d = d(\vec{X})$, $\vec{n} = const$, which corresponds to the equation (36). In our view, it is a more realistic situation, as in the SmA the discrete frequency exists in the direction perpendicular to the layers. It follows that the force must be directed along the director (36), rather than perpendicular to it (in case when $\vec{n} = \vec{n}(\vec{X})$, $d = const$ [2] the force is directed along $\delta\vec{n}$, namely parallel to the layers at small deformations [25]).

## 7. Ginzburg-Landau model and electron-phonon interaction.

By analogy with the spatial translations let us consider the model with local transformational properties of the OP at temporal translations $\hat{\tau} \in T$: $\hat{\tau}\psi_\omega = e^{i\omega(x)\tau}\psi_\omega$, $\hat{\tau}\psi_{-\omega} = e^{-i\omega(x)\tau}\psi_{-\omega}$ (2). Here, $T$ is a group of temporal translations and $\omega$ is the frequency that is the parameter of the IR of the temporal translations group. In this case the extended derivative has the form:

$$D_j \psi_\omega = \left(\frac{\partial}{\partial X_j} - i\gamma A_j^\omega\right)\psi_\omega, \quad D_j \psi_{-\omega} = \left(\frac{\partial}{\partial X_j} - i\gamma A_j^{-\omega}\right)\psi_{-\omega}, \quad (38)$$

where $\gamma$ is phenomenological charge, and the compensating fields $A_j^\omega$ and $A_j^{-\omega}$ are transformed under the operator $\hat{\tau}$ as:

$$\hat{\tau}(\gamma A_j^\omega) = \gamma A_j^\omega + \frac{\partial \omega}{\partial X_j}\tau, \quad \hat{\tau}(\gamma A_j^{-\omega}) = \gamma A_j^{-\omega} - \frac{\partial \omega}{\partial X_j}\tau. \quad (39)$$

As in (7) from two fields $A_j^\omega$ and $A_j^{-\omega}$ we can get to one independent field $A_j^\omega \equiv A_j^\Omega$, because $A_j^{-\omega} = -A_j^\omega = -A_j^\Omega$ (38,39). Then the extended derivative $D_j \psi_{-\omega}$ in (38) will look like:

$D_j \psi_{-\omega} = \left(\frac{\partial}{\partial X_j} + i\gamma A_j^\Omega\right)\psi_{-\omega}$. Under the effect of the point group operations the field $A_j^\Omega$ is transformed as $\partial\omega/\partial X_j$ (39) and, therefore, changes its sign upon inversion of time $\hat{R}$:

$$\hat{R}(A_j^\Omega) = -A_j^\Omega, \quad \hat{R}(\psi_\omega) = \psi_{-\omega}. \quad (40)$$

Note that in the electrodynamics field theory [3,9] transformation of electromagnetic potential under temporal inversions: $\hat{R}(A_j) = -A_j$, had to be postulated because the gauge parameter $\alpha$ in (15) characterises the internal symmetry U(1) and is invariant under temporal inversions.

We can construct a local phenomenological Landau potential which is invariant with respect to elementary temporal translations (without taking anisotropy into account):

$$F = a\psi_\omega\psi_{-\omega} + \frac{b}{2}(\psi_\omega\psi_{-\omega})^2 + \beta(\vec{\nabla} - i\gamma\vec{A}^\Omega)\psi_\omega(\vec{\nabla} + i\gamma\vec{A}^\Omega)\psi_{-\omega} + F_e(\vec{\nabla}\times\vec{A}^\Omega). \quad (41)$$

Expression (41) is similar to the Ginzburg-Landau potential up to coefficients [18], if we assume that $\gamma \vec{A}^\Omega = \frac{2e}{\hbar c}\vec{A}$, $\psi_\omega = \psi$ and $\psi_{-\omega} = \tilde{\psi}$:

$$F = a|\psi|^2 + \frac{b}{2}|\psi|^4 + \frac{\hbar^2}{4m}\left|(\vec{\nabla} - \frac{2ie}{\hbar c}\vec{A})\psi\right|^2 + \frac{1}{8\pi}\vec{B}^2, \qquad (42)$$

Here $\vec{B} = \vec{\nabla} \times \vec{A}$ is a magnetic field.

In reality, the phenomenological charge $\gamma$ is related to the physical constants with expression:

$$\gamma A_j^\Omega = \frac{e}{\hbar c} A_j, \qquad (43)$$

where $e$ is the electron charge, $\hbar$ is the Planck constant, $c$ is the speed of light.

Expression (43) derives from the principle of least action for the charge $\gamma$ located in field $A_j^\Omega$. Action for the charge $\gamma$ in field $A_j^\Omega$ by analogy with (28) reads:

$$S = -\hbar\gamma \int_a^b (-A_j^\Omega dx^j + c\varphi^\Omega dt), \qquad (44)$$

where $\varphi^\Omega$ is a potential field that compensates the temporal derivatives of the OP with $\omega \neq 0$. As shown in par. 5, $\hbar\kappa_i = p_i$ is a momentum. At the same time, $\kappa_i$ is a phenomenological charge related to the law of conservation of momentum (12), that derives from the invariance of Landau potential in relation to elementary spatial translations (par. 3). Therefore, we can assume that $\hbar\gamma$ is energy. It is related to the law of conservation that derives from the invariance of the Landau potential in relation to the temporal translations (2,39). On the other hand, the action for electric charge in the electromagnetic field looks like this (27) [23]. If we compare it to (44), we have a connection between the phenomenological charge $\gamma$ and the compensating field $A_j^\Omega$, on the one hand, and the electric charge $e$ and the electromagnetic potential $A_j$, on the other hand:

$\hbar\gamma A_j^\Omega = \frac{e}{c} A_j$, which leads to $\gamma A_j^\Omega = \frac{e}{\hbar c} A_j$ (43). Since the dimentionality of the phenomenological charge $\gamma$ - $[1/c]$, then the physical dimensionality of compensating field $A_j^\Omega$ equals to $[c/м]$ (38,39). Dimensionality $A_j^\Omega$ is not the same as the dimensionality of the electromagnetic potential $A_j$. However, using expression (43), it is possible to replace $\gamma A_j^\Omega$ in the extended derivative (41) with $\frac{e}{\hbar c} A_j$. This proves that electromagnetic potential $A_j$ is the compensating field for the OP with local transformational properties in temporal translations (2). This proof is necessary here because generally speaking the OP is not a quantum mechanics wave function (42).

The obtained extended derivative differs from the extended derivative of the potential (42) by multiplier two. If the choice of the coefficients in front of derivative in the expression (42) is conditional [18], then the charge doubling in the extended derivative of the Ginzburg-Landau model [19] is necessary to get a correct expression for the density in the London equations and the quantum of the magnetic flux. In potential (41) that was constructed for model $\vec{k} = 0$, we decided not to double the phenomenological charge $\gamma$ (38,39) artificially, as it is not needed to obtain the correct expression of magnetic flux quantum in the model with $\vec{k} \neq 0$, see par. 8.

As is known, the superconducting states in the external magnetic field are inhomogeneous and have defects in the form of dislocations (see reviews [16,17] and references specified there). Therefore, it is possible that the superconducting OP has local transformational properties in relation to the spacial translations subgroup (3). Hence, to describe the inhomogeneous state of the superconductor a local Landau potential for OP needs to be constructed with $\omega = \omega(\vec{X})$ and $\vec{k} = \vec{k}(\vec{X})$ (2,3). In this case, the extended derivative will include linear terms of both elastic distortion tensor $A_{pj}$ (phonon potential) and electromagnetic vector-potential $A_j$. The existence of such a derivative and its linear form is evident from (31), in transition to the Hamiltonian formalism [23].

The main question to be answered is: is there any set of equations of state for the local Landau potential that takes into account compensating phonon field $A_{pj}$ and electromagnetic potential $A_j$ in the extended derivative, solutions that describe the Meissner effect? The matter is that if the elastic degrees of freedom in the form of derivatives from the displacement vector (the change in the displacement vector is usually related to phonons - lattice oscillations) are taken into account in the Ginzburg-Landau potential, it is not possible to obtain London equations from the set of equations of state. Indeed, the interaction between the OP and the deformation tensor, e. g. $c_{pqij}|\psi|^2 u_{pi} u_{qj}$, in the Ginzburg-Landau potential would lead to the OP module being dependent on the displacement vector $u_i$, which, in its turn, depends on $\vec{X}$, by definition. Hence, approximation $|\psi| = const$ cannot be used for this potential and, therefore, London equations cannot be obtained [18]. Apparently, it is due to this circumstance that there is no analogue for electron-phonon interaction in the Ginzburg-Landau potential (42). On the other hand, as is known [20], the superconducting gap opens as a result of the electron-phonon interaction. Thus, in the microscopic theory electron-phonon interactions are necessary [20] to describe the superconductivity, but in the phenomenological theory taking into account of elastic variables in the nonequilibrium potential does not provide the correct result. It would appear that the matter is in the way phonons are described. It was shown above that describing phonons via distortion tensor leads to the correct definition of the sound velocity (par. 4) and agrees with quantum-mechanic wave representations (par. 5). Furthermore, the minimal interaction between the distortion tensor $A_{pj}$ and the OP formally is analogous to the electromagnetic interaction and corresponds to electron-phonon interactions in the microscopic theory [18-20]. At the same time, the description of phonons through the displacement vector in the inhomogeneous state with dislocations is unlikely as the displacement vector is not determined, generally speaking, in the state with dislocations.

So, let us construct the OP for the superconducting state. For this we chose the IR with $\vec{k} \neq 0$: $\psi_{\vec{k}_l} = \psi_{\vec{k}_l(\vec{X})}$, and require that it has local transformational properties in relation to the spatial translations subgroup:

$$\hat{\tau}\psi_{\vec{k}_l,\omega} = e^{i\omega(x)\tau}\psi_{\vec{k}_l,\omega}, \quad \hat{\tau}\psi_{\vec{k}_l,-\omega} = e^{-i\omega(x)\tau}\psi_{\vec{k}_l,-\omega}. \tag{45}$$

The inversion of time acts as $\hat{R}(\psi_{\vec{k}_l,\omega}) = \psi_{\vec{k}_l,-\omega}$ for each $\vec{k}_l$. Taking into account that nontrivial transformations of the OP at temporary translations (45) doubles the dimensionality of the IR with $\vec{k} \neq 0$: $\psi_{\vec{k}_l,\omega}$, $\psi_{\vec{k}_l,-\omega}$ (note, $\vec{k}_l$ and $-\vec{k}_l$ are different vectors in star $\{\vec{k}\}$). The quadratic invariant in this case has the form:

$$I = \sum_l \psi_{\vec{k}_l,\omega}\psi_{-\vec{k}_l,-\omega} + \psi_{\vec{k}_l,-\omega}\psi_{-\vec{k}_l,\omega}. \tag{46}$$

The extended derivatives for this model according to (5,38):

$$D_j\psi_{\vec{k}_l,\omega} = \left(\frac{\partial}{\partial X_j} - i\sum_p \kappa_p A^l_{pj} - i\gamma A^\Omega_j\right)\psi_{\vec{k}_l,\omega}, \quad D_j\psi_{\vec{k}_l,-\omega} = \left(\frac{\partial}{\partial X_j} - i\sum_p \kappa_p A^l_{pj} + i\gamma A^\Omega_j\right)\psi_{\vec{k}_l,-\omega},$$

(47)

$$D_j\psi_{-\vec{k}_l,\omega} = \left(\frac{\partial}{\partial X_j} + i\sum_p \kappa_p A^l_{pj} - i\gamma A^\Omega_j\right)\psi_{-\vec{k}_l,\omega}, \quad D_j\psi_{-\vec{k}_l,-\omega} = \left(\frac{\partial}{\partial X_j} + i\sum_p \kappa_p A^l_{pj} + i\gamma A^\Omega_j\right)\psi_{-\vec{k}_l,-\omega}.$$

Quadratic gradient invariants, which are responsible for minimal interaction, may be written in the form:

$$F_i = \beta \sum_{jl} (D_j\psi_{\vec{k}_l,\omega} D_j\psi_{-\vec{k}_l,-\omega} + D_j\psi_{\vec{k}_l,-\omega} D_j\psi_{-\vec{k}_l,\omega}). \tag{48}$$

Here $\beta$ is a scalar. As anisotropy does not affect the conclusions below, expressions $I$ and $F_i$ (46,48) do not take into account anisotropy.

For model (45) the local Landau potential has the form:

$$F = F_L + F_i + F_e(\vec{\nabla}\times\vec{A}) + F_f(-e_{ijk}\partial A_{pk}/\partial X_j), \tag{49}$$

where $F_L = aI + b/2 I^2$ is the conventional Landau potential depending only on the OP components, $F_i$ is a part of the potential containing the OP derivatives and is responsible for the minimal interaction, $F_e(\vec{\nabla}\times\vec{A})$ and $F_f(-e_{ijk}\partial A_{pk}/\partial X_j)$ are quadratic forms corresponding to the free electromagnetic and the elastic energy respectively. Maxwell equations of state (49) have the form:

$$\delta F/\delta\vec{A} = \partial F_i/\partial\vec{A} - \vec{\nabla}\times(\partial F_e/\partial(\vec{\nabla}\times\vec{A})) = 0. \tag{50}$$

Let us look for solutions to the set of equations of state when all $\psi_{\vec{k}_l,\omega}$ are equal to zero except for four components: $\psi_{\vec{k},\omega}$, $\psi_{\vec{k},-\omega}$, $\psi_{-\vec{k},\omega}$, $\psi_{-\vec{k},-\omega}$, with some designated vectors $\vec{k}$ and $-\vec{k}$. We should note that equations of state always have such solutions [25]. Vector potential $A_j$ and distortion tensor $A_{pj}$ in an explicit form without derivatives only occur in $F_i$, that is why the vector of current $\vec{j} = \partial F_i/\partial\vec{A}$ contains only terms that are linear in $A_j$ with coefficients $\psi_{\vec{k},\omega}\psi_{-\vec{k},-\omega} + \psi_{\vec{k},-\omega}\psi_{-\vec{k},\omega}$, and the tensor components $A_{pj}$ occur in the expression for the current with coefficients: $\psi_{\vec{k},\omega}\psi_{-\vec{k},-\omega} - \psi_{\vec{k},-\omega}\psi_{-\vec{k},\omega}$ (47,48).

In general, all four components - $\psi_{\vec{k},\omega}$, $\psi_{\vec{k},-\omega}$, $\psi_{-\vec{k},\omega}$, $\psi_{-\vec{k},-\omega}$ are independent. Let us look for a solution to the set of equations of state in this form:

$$\psi_{\vec{k},\omega} = \rho e^{i\phi_1}, \quad \psi_{-\vec{k},-\omega} = \rho e^{-i\phi_1}, \quad \psi_{-\vec{k},\omega} = \rho e^{i\phi_2}, \quad \psi_{\vec{k},-\omega} = \rho e^{-i\phi_2}. \tag{51}$$

Solution (51) is the simplest which reflects the OP non-trivial transformational properties with respect to the both temporal and spatial inversions. When the OP corresponds to (51), coefficient, composed of the OP components, vanishes in front of the phonon potential $A_{pj}$ in the expression for the current: $\psi_{\vec{k},\omega}\psi_{-\vec{k},-\omega} - \psi_{\vec{k},-\omega}\psi_{-\vec{k},\omega} = \rho^2 - \rho^2 = 0$, and is doubled in front of the electromagnetic potential $A_j$: $\psi_{\vec{k},\omega}\psi_{-\vec{k},-\omega} + \psi_{\vec{k},-\omega}\psi_{-\vec{k},\omega} = 2\rho^2$. As the phonon potential $A_{pj}$ is not part of the expression for the current for solutions (51), so the electromagnetic potential $A_j$ vanishes in the expression for stress tensor $\sigma_{pj} = \partial F_i/\partial A_{pj}$. According to (43,47,48), when all the OP components are equal to zero except for four components with the designated vectors $\vec{k}$ and $-\vec{k}$, the current density is:

$$\vec{j} = -i\beta \frac{e}{\hbar c}(\psi_{-k,-\omega}\vec{\nabla}\psi_{k,\omega} - \psi_{k,\omega}\vec{\nabla}\psi_{-k,-\omega} + \psi_{k,-\omega}\vec{\nabla}\psi_{-k,\omega} - \psi_{-k,\omega}\vec{\nabla}\psi_{k,-\omega}) -$$

$$2\beta \frac{e^2}{\hbar^2 c^2}(\psi_{\vec{k},\omega}\psi_{-\vec{k},-\omega} + \psi_{\vec{k},-\omega}\psi_{-\vec{k},\omega})\vec{A} . \qquad (52)$$

For solution (51) from (52) we obtain:

$$\vec{j} = 2\beta \frac{e}{\hbar c}\rho^2(\vec{\nabla}\phi_1 + \vec{\nabla}\phi_2 - 2\frac{e}{\hbar c}\vec{A}) . \qquad (53)$$

London equations result from (50,53).

Thus, the expression for the superconducting current (52,53) does not explicitly depend on the distortion tensor and internal stresses and describes the Meissner effect. Furthermore, for solution (51), the coefficient in front of the electromagnetic potential is doubled in the expression for the current (53), without doubling of the phenomenological charge in the extended derivative (47). The necessity for this doubling was proven in the paper by L.P. Gorkov [19] and is related to the quantum-mechanical description of the superconductivity.

**8. Minimal magnetic flux**

Doubling of the coefficient in front of the electromagnetic potential in the Ginzburg-Landau theory was necessary not only in order to obtain correct London equations but also to calculate the minimal magnetic flux [18]. This is a fundamental question. Let us demonstrate that solution (51) for potential (49) provides the correct expression for a quantum of the magnetic flux without doubling phenomenological charge in the extended derivative.

As the OP interacts with the electromagnetic potential (43,47,48), then we can obtain the expression for the minimal magnetic flux from (53), by repeating the reasoning in [18] for the quantum-mechanical wave function. Therefore, when the current is equal to zero, the following relationship should be valid:

$$\oint_L (\vec{\nabla}\phi_1 + \vec{\nabla}\phi_2)d\vec{l} = 2\frac{e}{\hbar c}\oint_L \vec{A}d\vec{l} . \qquad (54)$$

From the OP phase uniqueness (51) and definition $\vec{\nabla}\times\vec{A} = \vec{B}$ it follows that:

$$2\pi q + 2\pi r = 2\frac{e}{\hbar c}\int_{S_L}\vec{B}d\vec{s} = 2\frac{e}{\hbar c}\Phi , \qquad (55)$$

where $q, r \in Z$, and $\Phi$ is the magnetic flux through the circuit $L$.

The minimal magnetic flux occurs when one of the two whole numbers $q, r$ equals 1, and the other equals 0:

$$\Phi_{min} = \frac{\pi \hbar c}{e}. \tag{56}$$

As is known, expression (56) is related to the definition of the fundamental constants and is the criterion for the correctness of the theory.

Let us demonstrate that solution $\phi_2 = \phi_1$ in (51) is equivalent to the Ginzburg-Landau model [1]. Indeed, in this case it follows from (51): $\psi_{\vec{k},\omega} = \psi_{-\vec{k},\omega}$, which is equivalent to model $\vec{k} = 0$. It is easy to see if an expression analogous to (53) is written for the stress tensor:

$$\sigma_{pj} = 2\beta \kappa_p \rho^2 (\frac{\partial}{\partial X_j} \phi_1 - \frac{\partial}{\partial X_j} \phi_2 - 2\kappa_p A_{pj}). \tag{57}$$

Under the condition $\phi_2 = \phi_1$ the expression for the stress tensor should be identically zero $\sigma_{pj} \equiv 0$. Since when $\phi_2 = \phi_1$ is inserted in (57) the derivatives vanish, the expression (57) is no longer invariant with respect to elementary translations.

A similar situation occurs when $\phi_2 = -\phi_1$. Then it follows form (51) that $\psi_{\vec{k},\omega} = \psi_{\vec{k},-\omega}$. This means that using OP components it is not possible to construct a combination that would change its sign upon time inversion, i.e. it is not possible to determine the density of the current. Whence follows that $\vec{j} \equiv 0$ in the state with $\phi_2 = -\phi_1$, $\omega = 0$ (53).

Thus, under the condition $\phi_2 = \phi_1$ in (51), potential (49) represents a doubled Ginzburg-Landau potential [1], where the OP does not interact with the phonon field ($\vec{k} = 0$). In this case the expression for the current (53) reads: $\vec{j} = 4\beta \frac{e}{\hbar c} \rho^2 (\vec{\nabla} \phi_1 - \frac{e}{\hbar c} \vec{A})$ and leads to doubling of the minimal magnetic flux (54-56): $4\pi q = 2 \frac{e}{\hbar c} \Phi'_{min}$. In order to obtain the correct result within [1] theory, doubling of the coefficient before the charge in the extended derivative had to be performed [18,19], the reason for this being electron pairing in a superconductive state [20]. As a result of this, there is an opinion that the phenomenological coefficients of the potential can be adjusted not only for the reasons of microscopy, but this step needs to be proven via the microscopic theory. It should be noted that electron pairing that is used to justify doubling of charge is easily observed in the model with $\vec{k} \neq 0$. Since as a solution the four OP components were chosen with two opposite vectors $\vec{k}$ and $-\vec{k}$ from star $\{\vec{k}\}$ they can be associated with two electrons with opposite momentums. In par. 5 we have proven that the transition from vector $\vec{k}$ IR to the wave vector and wave momentum is correct in the local Landau theory. Therefore, the problem of doubling of the interaction phenomenological charge in the Ginzburg-Landau's model is related to the choice of the IR with $\vec{k} = 0$. In our opinion, the microscopic proof of the phenomenology is not required because phenomenology is based on symmetry and experiment. It is obvious that phenomenological equations of Maxwell (50), Newton (12), Bernoulli (23) etc. do not result from microscopic representations.

Let us demonstrate that Burgers vector is the analogue of the minimal magnetic flux for the phonon potential $A_{pj}$. For this purpose we repeat the reasoning for the stress tensor (57) that was given above for the current density (53):

$$\oint_L (\vec{\nabla}\phi_1 - \vec{\nabla}\phi_2)d\vec{l} = 2\kappa_p \oint_L A_{pj} dl_j . \qquad (58)$$

Whence, according to Stokes' transformation for (11), we obtain:

$$2\pi q - 2\pi r = -2\kappa_p \int_{S_L} \rho_{pj} ds_j = -2\kappa_p B_p . \qquad (59)$$

$B_{p\min} = \dfrac{\pi}{\kappa_p} = \dfrac{\lambda_p}{2}$ is the minimal Burgers vector for states with $\phi_2 \neq -\phi_1$, with defined dynamic variables, equal to half wave length which agrees with the stationary waves concept.

For the state when $\psi_{\vec{k},\omega} = \psi_{\vec{k},-\omega}$ ($\phi_2 = -\phi_1$) in (51), the minimal Burgers vector is equal to $B_{p\min} = \dfrac{2\pi}{\kappa_p} = a_p$ of the elementary lattice period, q.e.d. This means that a minimum density flow of dislocations is an elementary period of the crystal lattice, and the phenomenological charge of the interaction of the OP and the phonon field is $\kappa_p = \dfrac{2\pi}{a_p}$. In contrast to the fundamental values of the magnetic flow quantum, the minimum Burgers vector is linked to the period of the crystal lattice. Determination of the minimum Burgers vector requires elementary period that binds definition of dislocations in a continuous medium with the definition of dislocations in the crystal lattice.

## 9. The gap d-symmetry for states with $\vec{k} \neq 0$.

In par. 7 and 8 we analysed the simplest non-trivial solution (51) that takes into account local transformational properties (2) as well as (3). It is obvious that a full analysis of solutions for the set of equations of state is required with respect to symmetry. Let us look for solutions for the four OP components (51), when all their phases are different:

$$\psi_{\vec{k},\omega} = \rho e^{i\phi_1}, \; \psi_{-\vec{k},\omega} = \rho e^{i\phi_2}, \; \psi_{\vec{k},-\omega} = \rho e^{-i(\phi_2 - \varepsilon_2)}, \; \psi_{-\vec{k},-\omega} = \rho e^{-i(\phi_1 - \varepsilon_1)}. \qquad (60)$$

Here $\varepsilon_1$, $\varepsilon_2$ characterise difference in phase between OP components, hence they are invariant in elementary translations. It is not possible to construct real invariant combinations similar to (46) when all phases $\phi_1, \phi_2, \varepsilon_1, \varepsilon_2$ are not equal to zero. In order to construct real non-equilibrium Landau potential for (60) the OP complex-conjugate components should be taken into account:

$$\tilde{\psi}_{\vec{k},\omega} = \rho e^{-i\phi_1}, \; \tilde{\psi}_{-\vec{k},\omega} = \rho e^{-i\phi_2}, \; \tilde{\psi}_{\vec{k},-\omega} = \rho e^{i(\phi_2 - \varepsilon_2)}, \; \tilde{\psi}_{-\vec{k},-\omega} = \rho e^{i(\phi_1 - \varepsilon_1)}. \qquad (61)$$

It is easy to see that generally space-time inversion is not equivalent to the complex conjugate for the OP components in (60,61). The quadratic invariant composed of components (60,61), will take the form:

$$I = \psi_{\vec{k},\omega}\tilde{\psi}_{\vec{k},\omega} + \psi_{-\vec{k},\omega}\tilde{\psi}_{-\vec{k},\omega} + \psi_{\vec{k},-\omega}\tilde{\psi}_{\vec{k},-\omega} + \psi_{-\vec{k},-\omega}\tilde{\psi}_{-\vec{k},-\omega} + \qquad (62)$$

$$+ \psi_{\vec{k},\omega}\psi_{-\vec{k},-\omega} + \tilde{\psi}_{\vec{k},\omega}\tilde{\psi}_{-\vec{k},-\omega} + \psi_{\vec{k},-\omega}\psi_{-\vec{k},\omega} + \tilde{\psi}_{\vec{k},-\omega}\tilde{\psi}_{-\vec{k},\omega}.$$

The first four terms and the last four terms in (62) are invariant in translations (2,3) and space-time inversion. Their sum in quadratic invariant (62) is imposed by the CPT symmetry of equations of state. It is easy to observe that with variation of $I$ (62) in relation to any OP component (60,61) equations of state do not change at simultaneous operation of time inversion, space inversion and complex conjugate (as known [9], complex conjugate procedure is associated with charge conjugation symmetry). If we insert (60,61) into (62), we obtain:

$$I = \rho^2(2 + 2 + 2\cos\varepsilon_1 + 2\cos\varepsilon_2) \tag{63}$$

Taking into account complex conjugate OP components (61) in $F_i$ (48), by analogy with (62), we have an expression for the current $\vec{j} = \partial F_i / \partial \vec{A}$:

$$\vec{j} = 2\beta \frac{e}{\hbar c} \rho^2 \{ [\vec{\nabla}\phi_1 + \vec{\nabla}(\phi_1 - \varepsilon_1) - 2\frac{e}{\hbar c}\vec{A}](1 + \cos\varepsilon_1) + [\vec{\nabla}\phi_2 + \vec{\nabla}(\phi_2 - \varepsilon_2)$$

$$-2\frac{e}{\hbar c}\vec{A}](1 + \cos\varepsilon_2) \}. \tag{64}$$

The quantum of magnetic flux can be calculated by equating both expressions in square brackets to zero. In the general case, when all phases in (60) differ, $\Phi_{\min}$ coincides with expression (56). In the case when one of $\varepsilon_1$, $\varepsilon_2$ in (60,61) is equal to zero, $\Phi_{\min}$ is also defined by expression (56). It is easy to verify that taking into account additional OP components, when there are more than two non-zero vectors $\vec{k}_l$, will not affect expression $\Phi_{\min}$ (56).

When $\varepsilon_1 = \varepsilon_2 = 0$ expression (60) becomes model (51). In this case the expression for the current (64) will be four times bigger than expression for the current (53), because it takes into account complex conjugate values (61).

For simplicity, let us consider solution (60) when $\varepsilon_1 = \varepsilon_2 = \varepsilon$. Then the expression for the current will be:

$$\vec{j} = 4\beta \frac{e}{\hbar c} \rho^2 [\vec{\nabla}\phi_1 + \vec{\nabla}\phi_2 - \vec{\nabla}\varepsilon - 2\frac{e}{\hbar c}\vec{A}](1 + \cos\varepsilon), \tag{65}$$

It follows from (60) that $\varepsilon$, in the general case, depends on the direction of vector $\vec{k}$: $\varepsilon = \varepsilon_{\vec{k}}$. Let us assume that at a certain position of vector $\vec{k}$, $\cos\varepsilon_{\vec{k}} = -1$. If the point group of the crystal has a four-fold axis, then upon the rotation in the reciprocal space to angle $\pi/2$ around this axis nothing should change by symmetry. This leads to a conclusion that phase $\varepsilon_{\vec{k}}$, in the simplest non-trivial case, has dependence: $\varepsilon_{\vec{k}} = 4\theta$, where $\theta$ is the angle of rotation around the four-fold axis. If we insert $\varepsilon_{\vec{k}} = 4\theta$ into (63), we obtain:

$$I = 4\rho^2 (1 + \cos 4\theta). \tag{66}$$

Expression (66) corresponds to the d-symmetry of the superconducting gap. In this case the d-symmetry of the superconducting gap is associated with the four-fold axis of symmetry and not with the d-electron pairing. Generally speaking, the IR of the continuous group of rotations cannot be used to describe the d-symmetry of the superconducting gap in a crystal with the four-fold axis of symmetry.

It should be noted that if the symmetry axis is two- or three-fold, then the multiplier before the rotation angle in (66) will be 2 or 3 respectively.

Thus, dependence (66) which characterises d-symmetry of the superconducting gap occurs in low-symmetry state that is described by solution (60,61). Here non-invariance of the OP components (60,61) under the influence of the CPT operator should not be confused with the invariance of equations of state upon the CPT transformation that results from (62). For solutions (51) the space-time inversion is equivalent to the complex conjugate operation. Hence, we did not use complex conjugate IP components in par. 7 and 8; they were not necessary to construct the real Landau potential. In this respect solutions where the complex conjugate is equivalent to the space-time inversion can be singled out into a separate symmetry category. For such solutions quadratic invariant $I$ does not contain factor $(1+\cos\varepsilon_{\vec{k}})$, which is associated with zeroing of the superconducting gap in designated directions of the reciprocal space.

States with the d-symmetry of the superconducting gap are usually described as a violation of the point symmetry in a crystal, see review [26] and references specified there. For this purpose, switched to the Fourier transform of the wave function $g(\vec{k}) = \iiint_V \psi(\vec{x})e^{i\vec{k}\vec{x}}d\vec{x}$ and expanded it by the IR of the group $\Im$. In [26] considered symmetry group: $\Im = U(1) \times R \times G$, where $U(1)$ is the internal gauge symmetry, $R$ is the time reversal symmetry and $G$ is the crystal point symmetry group. In this process the translational symmetry of the crystal is not taken into account, the translational subgroup does not make part of symmetry group $G$. This is due to the fact that "there is no additional spatial modulation in transition to the superconducting state". Hence, in [26] the OP is transformed as a combination of components of vector $\vec{k}$ upon transformations from the crystal point symmetry group.

Generally speaking, the translation subgroup in the Landau theory can be disregarded when the symmetry of the high-symmetry phase does not have it, i.e. when there is no long-range order in the high-symmetry phase in the first place. In the case presented in [26] the long-range order is assumed to be present in the crystal high-symmetry phase. It is obvious that the nonequilibrium potential studied in [26] is invariant with respect to the translation subgroup of the high-symmetry phase. Since the OP is explicitly dependent on vector $\vec{k}$ and $g(\vec{k})$ is transformed upon translations into $e^{i\vec{k}\vec{x}}$ for each $\vec{k}$, according to (1).

In our opinion, when constructing the Landau potential its invariance with respect to the translation subgroup cannot be disregarded. As opposed to the symmetry group studied in [26], in this paper we use the symmetry group that is a direct product of groups: $\Im = T \times R \times G_0$, where $T$ is the group of temporal translations, $R$ is the time inversion, and $G_0$ is the space group of symmetry of the high-symmetry phase. Apart from the crystal point symmetry subgroup, the space group of symmetry $G_0$ contains the subgroup of spatial translations for the period of the elementary cell of the crystal. Therefore, when constructing invariants of the nonequilibrium Landau potential, invariants of the translational subgroup for the local IR (3) were constructed first of all and then - invariants of the crystal point symmetry subgroup (46,48,62). The structure of the Landau potential obtained in this way differs from the potential studied in [26]. As it was demonstrated above, for the crystal with a four-fold axis of symmetry the state with superconducting gap d-symmetry is described with the OP with $\vec{k} \neq 0$ and $\omega \neq 0$, components of which do not convert into complex conjugate components upon the space-time inversion.

It should be noted that solutions with two different phases in OP components can be obtained for the IR of the temporal translation group (2) with $\omega \neq 0$ and $\vec{k} = 0$, where $\Phi_{\min}$ will correspond to (56). This solution where the time inversion is not equivalent to the complex conjugate for the OP components: $\phi_2 = \phi_1 = \phi$, $\varepsilon_2 = \varepsilon_1 = \varepsilon$ in (60,61). However, in this solution phase difference $\varepsilon$ is not dependent on $\vec{k}$. In addition, in this case there are no solutions that would correspond to the conventional superconductivity: without factor $(1 + \cos\varepsilon)$ in $I$ and with correct $\Phi_{\min}$ (56). It also leads to a conclusion that to describe the superconductivity the IR needs to be considered with $\vec{k} \neq 0$ and $\omega \neq 0$. Therefore, the case with one phase: $\phi_2 = \phi_1 = \phi$, $\varepsilon_2 = \varepsilon_1 = 0$ in (60) ($\vec{k} = 0$) that corresponds to the Ginzburg-Landau model and results in doubling of the minimal magnetic flux can be considered degenerate. This is because the OP in the Ginzburg-Landau potential [1] is represented by the wave function with one phase by definition. In this sense, the model described in [26], with the group $U(1)$ does not differ from the Ginzburg-Landau theory, there must also be doubling of the phenomenological charge of interactions to obtain the correct quantum of the magnetic flow. The presented model with local transformational properties of the OP with respect to the translational subgroup always has solutions with two phases of the OP components since operations from the point group always make part of the space group. One element of the point symmetry group would be enough to make the IR of the translation subgroup two-dimensional. For example, the operation of reflection in a plane the normal vector of which coincides with the translation direction. Hence, complex representations considered here are two-dimensional as a minimum. This is a fundamental difference of the local Landau theory from the Abelian gauge model of the field theory U(1), with a single phase. Even if the non-Abelian gauge models contain multicomponent field complex functions, they operate with abstract internal symmetries that are not related to the space-time symmetry in any way [3]. For this reason they cannot be equivalent to states with $\vec{k} \neq 0$. Nonetheless, there are some attempts to obtain classical equations for physical quantities that were obtained in section 3,4 for a model with $\vec{k} \neq 0$, from the gauge field theory with abstract internal symmetries. In the next section we will look at one of them that relates to the gauge theory of dislocations.

**10. Kadić-Edelen model. The gauge theory of dislocations.**

Continuity equations in the form of the second Newton law (12) were for the first time ever obtained by Kadić and Edelen in the gauge theory of dislocations [12]. In this theory displacement vector $u_i$ is the main variable and the distortion tensor is compensating field $A'_{ij}$. Let us follow the derivation the Newton equations (12) in the gauge theory of dislocations in its modern interpretation [14,27].

As is known, in translations $b_i$ displacement vector $u_i$ does not change $\hat{b}_i(u_i) = u_i$, by definition [10]. The basis for the modern gauge theory of dislocations is the postulate on the locality of the space translations subgroup $b_i = b_i(X_i, T)$ in space $\{\vec{X}\}$, in which the displacement vector takes on its values:

$$\hat{b}_i(u_i) = u_i + b_i(\vec{X}, T). \tag{67}$$

In this case spatial and temporal derivatives of displacement vector $\partial u_i / \partial X_j$, $\partial u_i / \partial T$ become invariant with respect to the local translation subgroup. As according to (67) they are defined accurate to: $\partial b_i / \partial X_j$, $\partial b_i / \partial T$.

In order to construct the Lagrangian invariant under transformations (67), additional compensating fields are introduced into the theory - distortion compensating field $A'_{ij}$ and velocity compensating field $\upsilon'_i$, that are transformed in the following way:

$$\hat{b}_i(A'_{ij}) = A'_{ij} - \partial b_i/\partial X_j, \quad \hat{b}_i(\upsilon'_i) = \upsilon'_i - \partial b_i/\partial T. \tag{68}$$

Then, by replacing derivatives $\partial u_i/\partial X_j$, $\partial u_i/\partial T$ with translation-invariant combinations:

$$D_j u_i = \partial u_i/\partial X_j + A'_{ij}, \quad D_0 u_i = \partial u_i/\partial T + \upsilon'_i, \tag{69}$$

and by taking into account the proper invariants of the distortion tensor and velocity vector (68):

$$\rho_{ij} = e_{jkn} \partial A'_{in}/\partial X_k, \tag{70}$$

$$\varepsilon_{ij} = \partial \upsilon'_i/X_j - \partial A'_{ij}/\partial T. \tag{71}$$

From (11,17) and (70,71) $A'_{ij} = -A_{ij}, \upsilon'_i = -\upsilon_i$.

We obtain the invariant Lagrangian:

$$L(X_i, T) = L(D_0 u_i, D_j u_i, \rho_{ij}, \varepsilon_{ij}), \tag{72}$$

where $\rho_{ij}$ is the dislocation density [14], $Y_{ij}$ is the gradient invariant. The Lagrangian (72) is a quadratic form:

$$L(D_0 u_i, D_j u_i) = K(D_0 u_i) - \Pi(D_j u_i), \tag{73}$$

and is the difference between kinetic and potential energy. Furthermore, potential energy in (73) is the conventional potential in the theory of elasticity [10] where $\partial u_i/\partial X_j$ was substituted with $D_j u_i$ [14,27].

It follows from (69,72,73) that $\partial L/\partial A'_{ij} = -\sigma_{ij}$, $\partial L/\partial \upsilon'_i = p_i$. Then it is easy to derive continuity equations from equations of state: $\partial \sigma_{ij}/\partial X_j = \partial p_i/\partial T$ (see conclusion in section 3).

It is obvious that from the quadratic Lagrangian (72) linear equations of state can be obtained. As is known, they have wave solutions under an additional gauge condition for compensating fields. In this case this condition is pseudo-Lorentzian [27]:

$$\partial A'_{ij}/\partial X_j = \frac{1}{c_w^2} \partial \upsilon'_i/\partial T, \tag{74}$$

as opposed to the Lorentz condition (19) for compensating fields of the local OP. The sign of the right part of the expression (74) is determined by the type of the gradient invariant (71) (compare with the phonon field density (17)), which, in its turn, is determined by transformational properties (68) (compare with (6,14) and (15)).

It should be noted that the pseudo-Lorentzian gauge condition (74) does not occur here by chance. Compensating fields ($A'_{ij}, \upsilon'_i$) occur in (69) in the form of terms with displacement vector's derivatives ($\partial u_i/\partial X_j, \partial u_i/T$). So the conjugate values for the pair ($A'_{ij}, \upsilon'_i$) and ($\partial u_i/\partial X_j, \partial u_i/T$) are the same, it is stress tensor $\sigma_{ij}$ and momentum vector $p_i$ respectively. The transformation law (68) in translations for ($A_{ij}, \upsilon'_i$) is also defined from (67,69). It differs in principle from the transformation (6,14), so the translation invariant (71) differs from the density of the phonon field (17). In our opinion, this can explain the absence of the analogue for Coulomb force in the list of analogues [14], as it would have had the form: $p_i \partial \upsilon'_i/\partial X_j$, according to (71). As it was shown in par. 4 and proven in par. 5, the Coulomb force is equivalent to the potential member in the Euler hydrodynamics equations: $-p_i \partial \upsilon_i/\partial X_j$, which lead to the correct expression for the Bernoulli equation (23). The direction of the force here is of primary importance.

In fact, the difference between two approaches ([14] and the local Landau theory) consists not only in transformational properties upon translations for distortion 4-tensor (6,14) and (68), in signs of dislocation density (11) and (70), and analogues shown in the table in [14]. This difference is much more profound and we are going to discuss it below.

In their pioneering work Kadić and Edelen [12] obtained local transformations of the translation subgroup (67) from the Yang-Mills model that operates with abstract internal symmetries [3]. We will not attempt to repeat this derivation but will solve the problem of justification of local transformational properties (67) for displacement vector $u_i$ within a model with local translational symmetry of the crystal lattice.

As is known, the Landau theory of phase transitions describes transitions from high-symmetry state with a global symmetry group to low-symmetry state. On the other hand, if one assumes that it is group $a_i = a_i(\vec{X}, T)$ that is local, not representation parameters $k_i^l = k_i^l(\vec{X}, T)$ in (3,6,14), then the reasons why the local symmetry occurs in low-symmetry state can disregarded. In this case tensor $A_{ij}$ is defined accurate to $\partial a_i / \partial X_j$ according to (6) ($k_i^l = const$), and it follows from (14) that $\upsilon_i$ is defined accurate to $\partial a_i / \partial T$. Thus, variable compensating fields have the same dimension as the deformation tensor and the velocity vector. However, $A_{ij}$ and $\upsilon_i$ are defined as independent fields that cannot be represented in the form of space and time derivatives which are the deformation tensor and the velocity vector. Moreover, given the symmetry requirement in this model, the observables are asymmetrical combinations of space derivatives $A_{ij}$, that are defined accurate to the space derivatives of the displacement vector (11). As is known, due to this fact the distortion tensor [10] was introduced into the theory of dislocations.

On the other hand, in the classical theory of elasticity, where the principle variable is displacement vector $u_i$, the deformation tensor is an invariant of the translations subgroup in $\{\vec{X}\}$, by definition. Moving from continuous medium to the crystal lattice, in the Landau theory invariance of the deformation tensor will be required not only with respect to translations in $\{\vec{X}\}$, but also invariance with respect to translations for the elementary period of the crystal lattice. It is obvious that in the general case the situation with local subgroup of elementary translations $a_i = a_i(\vec{X})$ occurs in the deformed state. This arises from the very concept of the inhomogeneous deformation of the crystal lattice. Then, it follows from the definition of displacement vector $\vec{u}(\vec{X}) = \vec{X} - \vec{X}_0$ that:

$$\hat{a}_i(u_i) = u_i + a_i(\vec{X}) - a_i(\vec{X}_0), \tag{75}$$

where $a_i(\vec{X}_0)$ is the period of the lattice in non-deformed state, $a_i(\vec{X})$ is the period of the lattice after the deformation in point $\vec{X}$. According to (75) the space derivative of the displacement vector upon translation to the elementary period is transformed in the following way:

$$\hat{a}_i(\partial u_i / \partial X_j) = \partial u_i / \partial X_j + \partial a_i / \partial X_j \tag{76}$$

Therefore, the deformation tensor is not an invariant of the elementary translations subgroup when the lattice is taken into account. That is to say that the deformation tensor is not an observed value when the proximity of the continuous medium is not affecting it except for homogeneous deformations of the lattice. Hence, in the case of the local translational lattice symmetry we should be describing elastic properties of the lattice through distortion tensor $A_{ij}$ that fulfils the role of the compensating field (5,6). Furthermore, using components $A_{ij}$ we can construct invariants of the translations subgroup that would correspond to the dislocation density (11). According to (76) $\partial u_i / \partial X_j$ can be used as an OP compensating field (3,4), but it is not possible to compose antisymmetric invariants of the dislocation density (11) which occur with necessity in this model.

In essence, the Kadić-Edelen model postulates local transformations of the translations subgroup for the displacement vector (67). As we have just demonstrated, this postulate is justified for the inhomogeneous model with the crystal lattice. But the conclusion made in [12], in our opinion, is wrong. It would be better to go from deformation tensor $u_{ij}$ to distortion tensor $A_{ij}$ and use it as an OP compensating field. This is what was done with electromagnetic potential $A_j$ that was derived from Maxwell equations and then used in the field theory of electrodynamics as a compensating field for a complex wave function. However, in [12,14] the authors took a different approach and compensated the very displacements that in the state with dislocations are not defined, generally speaking. Having found that the derivative of displacement vector (67) and distortion tensor (11) are defined with accuracy to a gradient of the vector function, Kadić and Edelen introduced "extended derivatives" (69) that do not contain elongation of the derivative itself, for example (5,7,38,47). By analogy with electrodynamics, or because the initial formulation of equations was based on the Yang-Mills model, theory [12] became to be known as the gauge theory of dislocations. As is known, the gauge symmetry group is associated with the gauge interaction that is recorded via an extended derivative. However, there is no gauge interaction in the gauge theory of dislocations [12,14,27] as such. Kadić and Edelen essentially constructed a linear model where the "extended derivative" (69) does not contain an interaction, as opposed to (5). Here the point is not in the degree of Lagrangian invariants (72) but in the mechanism of constructing extended derivative. In the "extended derivative" itself (69) there is no minimal interaction analogous to, for example, $\kappa_p A_{pj}^l \eta_l$ (5) or $\gamma A_j^\omega \psi_\omega$ (38), and there is no interaction charge.

Furthermore, in the gauge theory of dislocations [12,14,27] the concept of the free distortion field $A_{ij}$, similar to the free electromagnetic field $A_j$, has no sense, since equations of state do not allow invariant solutions in the absence of displacement vector $u_i$. Indeed, equations of state for the Lagrangian (72) have the following form:

$$\frac{\partial L}{\partial A'_{ij}} = e_{jmn} \frac{\partial}{\partial X_m} \frac{\partial L}{\partial \rho_{in}} - \frac{\partial}{\partial T} \frac{\partial L}{\partial \varepsilon_{ij}} , \qquad (77)$$

where $\partial L / \partial A'_{ij} = -\partial \Pi / \partial A'_{ij}$ (73). When $u_i = 0$ (and consequently $\partial u_i / \partial X_j = 0$) the left part of the equation (77) does not vanish, as it happens in gauge models [3] where in the absence of the field function the current density equals to zero (52), by definition. According to (69,73), under condition $\partial u_i / \partial X_j = 0$ the left part (77) is a linear expression dependent only on components of distortion tensor $A'_{ij}$. That is why the left part (77) is non-invariant in translations (68,69), as opposed to its right part that contains invariants of the local translations subgroup (70,71). This implies that in the Kadić-Edelen model we can only talk about solutions that make corrections to displacement vector $u_i$ taking into account defects described by distortion tensor $A'_{ij}$. So phonons were associated in [27] only with the displacement vector waves, not with waves of free distortion 4-tensor ($A'_{ij}$, $\upsilon'_i$) under the condition (74).

Therefore, in this theory [12,14]: 1) not in all expressions constructed by analogy with electrodynamics the signs coincide with classical equations of the theory of elasticity and hydrodynamics; 2) there is no interaction in «extended derivative» (69); 3) there are no solutions for a free compensating field ($A'_{ij}$, $v'_i$), density of which corresponds to the sound. Nonetheless, it is difficult to overestimate this work by Kadić and Edelen. They were the first to define the compensating field in [12] as distortion tensor and to demonstrate that Newton equations (12) are continuity equations in the model with the local translational symmetry, analogous to continuity equations for the electric charge in the field theory of electrodynamics.

**11. Conclusion**

As it was demonstrated in [4,8] the minimal interaction (5,38) in the Landau theory derives from the local transformational properties of the OP (2,3) with respect to space-time translations subgroup. So the main mission of the local Landau theory was reduced to physical interpretation of the compensating field in the extended derivative. In par. 3 it was proven that the compensating field of the OP for the IR with $\vec{k} \neq 0$ is distortion tensor $A_{pj}$.

In papers [1,2,12,14], on the contrary, the minimal interaction was constructed on the basis of the interpretation of the compensating field: as electromagnetic potential in [1], director variation in [2] and distortion tensor in [12,14]. The choice of the interpretation of the compensating field determined the physical model. In the same way that the choice of the electromagnetic potential as a compensating field in the field theory of electrodynamics predetermines the physical dimension of the current that is expressed in a wave function, the choice of the interpretation of the compensating field in the Landau theory predetermines physical interpretation of OP translational invariants that represent the first integral linked to the elementary translations subgroup.

The main methodic result of this article, in our opinion, is the interpretation of distortion 4-tensor ($A_{ij}$, $\upsilon_i$) as a phenomenological phonon potential. The main argument for such an interpretation is the construction of the analogue for the electron-phonon interaction of the BCS theory [20] in the local Landau theory. This interpretation is also favoured by the obtained exact wave solutions for equations of state for 4-distortion free field that can be interpreted as sound waves. It should be noted that exact wave solutions for sound are not present in the Euler nonlinear equations of hydrodynamics [22] (if the sound were described by nonlinear equations, then the frequency of sound would have to change with the amplitude, which does not happen). Robustness criteria for the constructed formal description would be a correct expression of the quantum of the magnetic flux $\Phi_{\min}$ (56) without artificial doubling of phenomenological charge of the interaction.

The practical use of the theory with local transformational properties of the OP is related to the description of the superconducting gap d-symmetry in the low-symmetry state. This state calls for low-symmetry solutions (60) for which the space-time inversion is not equivalent to complex conjugate of the OP components (61). These solutions were not considered in the theory of superconductivity because the OP corresponding to the superconducting state usually had trivial transformational properties upon transformations from the translations subgroup: $\vec{k} = 0$ [1,18], or the translations subgroup was not taken into account in the construction of the Landau potential [26]. This paper studied states that have not only non-trivial properties in spatial translations: $\vec{k} \neq 0$, but also non-trivial properties in temporal translations: $\omega \neq 0$. In section 7 we demonstrated that the phenomenological model described by the OP with $\omega = \omega(\vec{X})$ contains the electromagnetic potential in the extended derivative as a compensating field and is equivalent to Ginzburg-Landau model [1]. Solution for the OP components with $\omega \neq 0$ и $\vec{k} \neq 0$ where space-time inversion is not equivalent to complex conjugate results in factor $(1+\cos\varepsilon_{\vec{k}})$ in superconducting gap $I$ (here $\varepsilon_{\vec{k}}$ is the difference of phases of the superconducting OP (60)). If the crystal has a four-fold axis of symmetry, this dependence has the form: $\varepsilon_{\vec{k}}=4\theta$. In the theory of superconductivity [18,26] only one phase of the OP was taken into account because model U(1) [3,9] was used. That is why low-symmetry solutions that lead to factor $(1+\cos\varepsilon_{\vec{k}})$ in quadratic invariant $I$ could not be formulated.

The use of the presented theory is also linked to additional elastic degrees of freedom in the form of distortion tensor $A_{pj}$ that occur in the nonequilibrium potential. Indeed, since the distortion tensor makes part of the OP extended derivative in a linear pattern, then the elasticity module is proportionate to coefficient $\beta$ before extended derivative (48) and to the OP module: $\partial\sigma_{pj}/\partial A_{pj}=4\beta\kappa_p^2\rho^2$ (57). The presence of additional elastic modules in the superconducting state, below the Curie second-order transition point, explains the non-linear anomaly of elastic properties of superconductors [28]. The anomaly of elastic properties of superconductors was usually explained by the interaction between the OP and deformation tensor $c_{pqij}|\psi|^2 u_{pi}u_{qj}$, where $|\psi|$ is the module of the wave function. However, as it was demonstrated above (par. 7), taking into account the deformation tensor in the Ginzburg-Landau potential does not allow to formulate London equations. Moreover, the deformation tensor is not invariant at elementary translations in inhomogeneous state of the crystal (par. 10) and for this reason it cannot be used to describe elastic properties of superconductors. As a matter of fact, when taking into account inhomogeneous deformations of the lattice in the superconducting state expression: $c_{pqij}|\psi|^2 u_{pi}u_{qj}$ transforms into corresponding translation-invariant expression (48), which is recorded through distortion tensor $A_{pj}$. It is possible to extract from (48) the analogue of elastic interaction between the OP and the deformation tensor that has the form: $4\beta\kappa_p\kappa_q\rho^2 A_{pi}A_{qj}$ (for simplicity, in (48) scalar coefficient $\beta$ is used that does not take into account anisotropy).

We wish to draw attention to the fact that in the formulated local Landau theory there are no other minimal interactions except for those that occur in extended derivative (47). Indeed, since the IR are chosen from the expansion of the function into the Fourier integral (1), thereby the OP transformational properties are fixed with respect to the translations subgroup. Therefore, rotation and reflection operations from the point symmetry group translate one IR basis function into another one and amount to orthogonal transformations of vectors $\vec{k}^l$ in the reciprocal space. Even if the rotation subgroup is infinite as, for example, in the case of deformed SmA, then this will lead to the IR of infinite dimensions (star of vectors $\vec{k}^l(\vec{X})$ is a cone in this case) but not to an additional continuous parameter of the IR that could be localised. Therefore, in the local Landau theory additional compensating fields will not occur in the extended derivative. In terms of construction, compensating fields are related to the IR parameters of the translations subgroup - $\vec{k}^l$, and not the rotations subgroup.

Distortion tensor interpretation as a phenomenological phonon potential, analogous to the electromagnetic potential, leads to unambiguous correspondence between the continual theory of dislocations and electrodynamics. A suggestion put forward in the review [13] that centrally symmetric Coulomb force corresponds to Peach-Koehler antisymmetric force, Lorentz force to force acting on moving dislocation [10], charge conservation law to "Burgers vector conservation law" does not stand up under scrutiny. In par. 4 it was demonstrated that 1) Coulomb force corresponds to a potential term in Euler equations of hydrodynamics (21) that leads to Bernoulli equation (23); 2) Lorentz force corresponds to Peach-Koehler force (24) that leads to the expression of vortex term in Euler expressions [22]; 3) the charge conservation law in the form of continuity equation corresponds to the differential formulation of the law of conservation of momentum in the form of the second Newton law (12).

As for possible analogy between the local Landau theory and the GTR, it should be noted that the crystal lattice deformation with local translational symmetry can be described with local metrics $g_{ij}(\vec{X})$ in subspace $\{\vec{x}\}_{\vec{X}}$, characterised by macro-coordinate $\vec{X}$. But when differentiating in the macrospace $\{\vec{X}\}$ there is no reason to introduce local metrics, symmetry group $\{\vec{X}\}$ is global. Indeed, when formulating extended derivatives it is possible to formally get from changing coordinates of vector $\vec{k}(\vec{X}) = \mu_i(\vec{X})\vec{e}_i$ to changing value of reciprocal space basis vectors $\vec{k}(\vec{X}) = \mu_i\vec{e}_i(\vec{X})$ of subspace $\{\vec{x}\}_{\vec{X}}$, as it was demonstrated in par. 2 in construction of extended derivatives (7). Change in value of basis vectors $\vec{e}_i(\vec{X})$ in $\{\vec{x}\}_{\vec{X}}$ is described by reciprocal space metrics $g_{ij}(\vec{X})$ when only diagonal terms in metric tensor are not equal to zero (the case when the off-diagonal components of the metric tensor are nonzero is beyond the scope of this article.) The transition from the local translational properties of the condensate, which is described - $k_i = k_i(\vec{X})$, to the deformation of the subspace $\{\vec{x}\}_{\vec{X}}$, which is described by the local metric tensor $g_{ij}(\vec{X})$ - is analogous to the principle of equivalence for the model. Analytically, this results from $\vec{k}(\vec{X}) = \mu_i(\vec{X})\vec{e}_i = \mu_i\vec{e}_i(\vec{X}) = \mu_i g_{ii}(\vec{X})\vec{e}_i$. In accordance with the main provisions of the Landau theory it is necessary for the nonequilibrium potential to be translation-invariant in transformations (3), see (4). At the same time, it is not essential how to describe local translational symmetry of the lattice in subspaces $\{\vec{x}\}_{\vec{X}}$: via $\mu_i(\vec{X})$, $\vec{e}_i(\vec{X})$ or $g_{ij}(\vec{X})$. Therefore, the locality of the metrics tensor $g_{ij}(\vec{X})$ in $\{\vec{x}\}_{\vec{X}}$ leads to a local group of Landau potential (3,9) and compensating fields the OP extended derivative (5) and not to local metrics of macrospace $\{\vec{X}\}$ and the GTR model.

We arrived at this conclusion from separating coordinates into microscopic - $\vec{x}$ and macroscopic - $\vec{X}$. It should be noted that Noether's theorem for the local Landau theory also leads to invariance of macro-coordinates $\vec{X}$ in elementary translations that are set on subspace $\{\vec{x}\}_{\vec{X}}$:

$$\hat{\vec{a}}\vec{X} = \vec{X}. \qquad (78)$$

Indeed, continuity equations (12) can be obtained directly from the set of equations of state, see conclusion in par. 3. On the other hand, equations (12) follow from Noether's theorem upon transformation of functions according to (3) and coordinates according to (78). Therefore, coordinates must be transformed identically (78), which means the independence of macro-coordinates $\vec{X}$ upon translations of high-symmetry phase in subspace $\{\vec{x}\}_{\vec{X}}$ for lattice period $\hat{\vec{a}}$. As it was shown above, elastic properties of such a model must be described by the distortion compensating field $A_{pj}$. In the absence of the lattice or when it can be ignored, elastic properties are described within the model of continuous medium through deformation tensor $u_{pj}$ [10]. Note that here there is no contradiction with transformation (76) because par. 10 considered a lattice in space $\{\vec{X}\}$ and not in subspace $\{\vec{x}\}_{\vec{X}}$ as in [12, 14] there is no division of the space into microscopic and macroscopic.

Thus, the used mathematical model with variations calculus applies condition (78) to macro-coordinates $\vec{X}$ and with necessity requires their independence in translations in local subspace $\{\vec{x}\}_{\vec{X}}$. As a matter of fact, condition (78) is a mathematical expression of the principle of local homogeneity of the physical space [8], it allows to use variations calculus tool in the local Landau theory. In the introduction, it was noted that the OP dependence from coordinates is not possible in the classical Landau theory [5], by definition (1). In [8] it was suggested to theoretically divide the space into infinite number of local subspaces $\{\vec{x}\}_n$ that are numbered with macro-coordinates $\vec{X}_n$, where $n \in N$, and to construct in $\{\vec{x}\}_n$ the local Landau potential. It is obvious that the numbering of local subspaces $\{\vec{x}\}_n$ with macro-coordinates $\vec{X}_n$ does not change upon elementary translations in subspaces $\{\vec{x}\}_n$. Transition from numbering local subspaces $\{\vec{x}\}_n$, where $n \in N$, to a continuum of subspaces $\{\vec{x}\}_{\vec{X}}$, where $\vec{X} \in \{\vec{X}\}$, is the main concept of the theory. It allows constructing variations calculus in space $\{\vec{X}\}$ in the local Landau theory. Variations calculus in space $\{\vec{X}\}$ was used in the Lifshitz theory of phase transitions of second degree into inhomogeneous state [6,7] and then in the Landau theory with local transformational properties of OP (3) [4]. The obtained correct expressions for physical quantities (12,23,24,26,43,56, etc.) give ground to the conclusion that physical space can be formally presented in the form of splitting into infinite number of local subspaces $\{\vec{x}\}_{\vec{X}}$ in each point $\vec{X} \in \{\vec{X}\}$, where $\{\vec{x}\}_{\vec{X}}$ and $\{\vec{X}\}$ are Euclidian spaces. In such a case, invariants of elementary translations subgroup that correspond to observed physical quantities (for example, (10,11)), take their values in $\{\vec{x}\}_{\vec{X}}$, and $\vec{X}$ plays the role of a parameter. Thus, mathematical model $\{\vec{x}\}_{\vec{X}}$ reflects the property of the physical space.

We are to note that the model with local transformation properties of the OP with respect to the subgroup of translations in the Landau theory is not equivalent to the gauge field theory. Compensating field and OP in the local Landau are converted under the other representations of the space group and have other dimension (number of components) than compensating fields and field functions in the gauge field theory. This is illustrated by the example of electrodynamics, which was considered in paragraphes 7 to 9. In essence, the local Landau theory is an alternative to the gauge field theory. It contains the classification of the solutions of equations of state on the symmetry of space-time (p.7-9), in contrast to the gauge field theory, which operates with the abstract gauge groups of the "internal symmetries". The attempt to relate the gauge field theory with the symmetry of space-time, such as the gauge theory of dislocations [12], has failed (par. 10).

The present model is not equivalent to the general relativity one. Despite the fact that the states with the local translational symmetry can be described by the local metric tensor $\vec{k}(\vec{X}) = \mu_i g_{ii}(\vec{X})\vec{e}_i$, the construction of the nonequilibrium potential in the Landau formalism leads to a model with tensor compensating fields of distortion in the extended derivative (5,7), and not to the model with the Christoffel symbols [23]. Thus, compensating fields in the Landau theory are defined up to the derivatives of the metric tensor according to (6), while the Christoffel symbols are functions of the metric tensor.

For the same reason, the attempt to use Christoffel symbols as compensating fields in the extended derivative in the description of dislocations in graphene does not stand up to scrutiny [29]. The metric tensor is an observable value and for this reason it or its combinations cannot be used as a compensating field. It is easy to see that the extended derivative in [30] is not invariant in the gauge transformation. While describing dislocations in graphene it should be born in mind that in inhomogeneous deformations of graphene where local translational symmetry is preserved the spinor field will be transformed analogously (3). Then the distortion tensor will occur with necessity in the extended derivative.

The developed phenomenological model for the local OP with $\omega = \omega(\vec{X})$ and $\vec{k} = \vec{k}(\vec{X})$ is consistent with the quantum-mechanical description. The need to address simultaneously the local transformation properties (2) and (3) was proved in par. 7 in the construction of nonequilibrium superconducting potential. States described in this OP have local wave properties. On the other hand, the particle has local wave properties in quantum mechanics. Indeed, according to quantum mechanics, there are small areas of space, characterized by $\vec{X}$ in which the particle has wave properties characterized by the wave parameters $\omega$ and $\vec{k}$. Consequently, the wave parameters of the particle must vary with the coordinate $\vec{X}$, since otherwise the particle would be a wave, and there would be no area of its location in space.

Moreover, the principle of least action (par. 5) and the principle of minimal interaction of the OP with the compensating field (par. 2, 7) was obtained by the correct expression for the phenomenological charge in the extended derivative (41, 43). Expression (43) includes the Planck constant and is consistent with the fundamental constant - the quantum of magnetic flow (56). In contrast to the phenomenological Ginzburg-Landau theory, where the quantum-mechanical wave function is used as the OP, the model with the local OP describes a wide range of the symmetry conditions, since the components of the OP may have different phases. For example, the state with the d-symmetry of the superconducting gap (p. 9) corresponds to the solution with the various phases of the OP components (60).

Transformation of time reversal symmetry is taken into account in the group in the theory of phase transitions, and plays an important role, for example in the description of the magnetic structures. Similarly, the decomposition of the density of states (1) under the IR symmetry groups, except in a subgroup of spatial translations we should consider a subgroup of time translations (45). Then the model with local OP, which is converted under the local IR with $\omega = \omega(\vec{X},T)$ and $\vec{k} = \vec{k}(\vec{X},T)$ and agrees with the field theory of quantum mechanics. In this case:

1. Minimal interaction is a consequence of local wave properties of the OP (2 and 3), ie basic concepts of quantum mechanics.

2. Field interactions correspond to the compensating field of the OP. This field compensates for changes in group settings $\omega(\vec{X},T)$ and $\vec{k}(\vec{X},T)$ of the IR, and consists of four electromagnetic potential ($A_j, \varphi$) (p. 7) and four-dimentional field distortion ($A_{ij}, \upsilon_i$) (p. 3), which corresponds to the phonon potential (p. 4).

Thus, the Landau-Lifshitz heterogeneous model [6], for the OP with the local transformation properties (2 and 3), describes the state with the local wave properties. Changing the wave parameters of the physical condition with the coordinate is the cause of the minimal interactions, in particular - the electromagnetic ones.

The author wishes to acknowledge Yu. M. Gufan for invaluable discussions and support while writing the present paper.